\documentclass[]{aastex7}
\shorttitle{Tracing the light}
\shortauthors{Lei He et al.}
\usepackage{amsmath}
\usepackage{breakurl}
\usepackage{dashrule}
\begin{document}
\title{Tracing the light: Identification for the optical counterpart candidates of binary black-holes during O3}
\correspondingauthor{Lei He, Zhengyan Liu, Wen Zhao}

\author[0000-0001-7613-5815]{Lei He}
\affiliation{Department of Astronomy, University of Science and Technology of China, Hefei, Anhui 230026, China}
\affiliation{School of Astronomy and Space Sciences, University of Science and Technology of China, Hefei, Anhui 230026, China}
\email[show]{helei0831@mail.ustc.edu.cn}

\author[0000-0002-2242-1514]{Zhengyan Liu}
\affiliation{Department of Astronomy, University of Science and Technology of China, Hefei, Anhui 230026, China}
\affiliation{School of Astronomy and Space Sciences, University of Science and Technology of China, Hefei, Anhui 230026, China}
\email[show]{ustclzy@mail.ustc.edu.cn}

\author[0000-0001-9098-6800]{Rui Niu}
\affiliation{Department of Astronomy, University of Science and Technology of China, Hefei, Anhui 230026, China}
\affiliation{School of Astronomy and Space Sciences, University of Science and Technology of China, Hefei, Anhui 230026, China}
\email{nrui@ustc.edu.cn}

\author[0009-0002-0900-9901]{Bingzhou Gao}
\affiliation{Department of Astronomy, University of Science and Technology of China, Hefei, Anhui 230026, China}
\affiliation{School of Astronomy and Space Sciences, University of Science and Technology of China, Hefei, Anhui 230026, China}
\email{supernova_gbz@mail.ustc.edu.cn}

\author[0009-0002-5634-8842]{Mingshen Zhou}
\affiliation{Department of Astronomy, University of Science and Technology of China, Hefei, Anhui 230026, China}
\affiliation{School of Astronomy and Space Sciences, University of Science and Technology of China, Hefei, Anhui 230026, China}
\email{zmetassin@mail.ustc.edu.cn}

\author[0009-0000-5321-5775]{Purun Zou}
\affiliation{Department of Astronomy, University of Science and Technology of China, Hefei, Anhui 230026, China}
\affiliation{School of Astronomy and Space Sciences, University of Science and Technology of China, Hefei, Anhui 230026, China}
\email{zpr2004@mail.ustc.edu.cn}

\author[0000-0001-6223-840X]{Runduo Liang}
\affiliation{National Astronomical Observatories, Chinese Academy of Sciences, 20A Datun Road, Beijing 100101, China}
\email{liangrd@bao.ac.cn}

\author[0000-0002-1330-2329]{Wen Zhao}
\affiliation{Department of Astronomy, University of Science and Technology of China, Hefei, Anhui 230026, China}
\affiliation{School of Astronomy and Space Sciences, University of Science and Technology of China, Hefei, Anhui 230026, China}
\email[show]{wzhao7@ustc.edu.cn}

\author[0000-0002-7152-3621]{Ning Jiang}
\affiliation{Department of Astronomy, University of Science and Technology of China, Hefei, Anhui 230026, China}
\affiliation{School of Astronomy and Space Sciences, University of Science and Technology of China, Hefei, Anhui 230026, China}
\email{jnac@ustc.edu.cn}

\author[0000-0002-4223-2198]{Zhen-Yi Cai}
\affiliation{Department of Astronomy, University of Science and Technology of China, Hefei, Anhui 230026, China}
\affiliation{School of Astronomy and Space Sciences, University of Science and Technology of China, Hefei, Anhui 230026, China}
\email{zcai@ustc.edu.cn}

\author[0000-0002-7835-8585]{Zi-Gao Dai}
\affiliation{Department of Astronomy, University of Science and Technology of China, Hefei, Anhui 230026, China}
\affiliation{School of Astronomy and Space Sciences, University of Science and Technology of China, Hefei, Anhui 230026, China}
\email{daizg@ustc.edu.cn}

\author[0000-0002-7330-4756]{Ye-Fei Yuan}
\affiliation{Department of Astronomy, University of Science and Technology of China, Hefei, Anhui 230026, China}
\affiliation{School of Astronomy and Space Sciences, University of Science and Technology of China, Hefei, Anhui 230026, China}
\email{yfyuan@ustc.edu.cn}

\begin{abstract}
  The accretion disks of active galactic nuclei (AGN) are widely considered the ideal environments for binary black hole (BBH) mergers and the only plausible sites for their electromagnetic (EM) counterparts. \citet{grahamLightDarkSearching2023} identified seven AGN flares that are potentially associated with gravitational-wave (GW) events detected by the LIGO-Virgo-KAGRA (LVK) Collaboration during the third observing run. In this article, utilizing an additional three years of Zwicky Transient Facility (ZTF) public data after their discovery, we conduct an updated analysis and find that only three flares can be identified. By implementing a joint analysis of optical and GW data through a Bayesian framework, we find two flares exhibit a strong correlation with GW events, with no secondary flares observed in their host AGN up to 2024 October 31. Combining these two most robust associations, we derive a Hubble constant measurement of $H_{0}= 72.1^{+23.9}_{-23.1} \ \mathrm{km \ s^{-1} Mpc^{-1}}$ and incorporating the multi-messenger event GW170817 improves the precision to $H_{0}=73.5^{+9.8}_{-6.9} \ \mathrm{km \ s^{-1} Mpc^{-1}}$. Both results are consistent with existing measurements reported in the literature.
\end{abstract}

\section{Introduction}

During the third observing run (O3), the LIGO-Virgo-KAGRA (LVK) collaboration reported approximately 80 confirmed binary black hole (BBH) mergers in the local Universe \citep{abbottGWTC3CompactBinary2023, abbottGWTC21DeepExtended2024}. These events predominantly originate via two formation channels \citep{mapelliFormationChannelsSingle2021}: the isolated evolution of stellar binary systems \citep{deminkChemicallyHomogeneousEvolutionary2016} and dynamic formation within dense environments, such as globular clusters \citep{rodriguezBinaryBlackHole2016} or the accretion disks of Active Galactic Nuclei \citep[AGN,][]{mckernanRampressureStrippingKicked2019}. While these pathways explain many observed BBH properties, several massive BBH merger progenitors detected by LVK, which reside within the pair-instability mass gap \citep{woosleyPulsationalPairinstabilitySupernovae2017}, challenge isolated evolution models and highlight the unique role of hierarchical (i.e., dynamical) merger in producing systems with component masses exceeding theoretical limits.

In hierarchical mergers, remnant black holes (BHs) from previous coalescences accumulate mass through successive mergers \citep{fragioneRepeatedMergersEjection2020}. This process naturally generates BHs with masses spanning $50-130 M_{\odot}$, bridging the pair-instability mass gap. Within this framework, the binary components are expected to have dimensionless spins approaching 0.7, which is a characteristic value of merger remnants as dictated by angular momentum conservation principles \citep{gerosaHierarchicalMergersStellarmass2021}.

AGN accretion disks represent a promising location for hierarchical mergers. Crucially, gravitational recoil velocities from mergers \citep{varmaEvidenceLargeRecoil2022} are typically insufficient to eject remnants from the disk's deep potential well, enabling successive mergers to build up BH masses. Coupled with the efficient binary formation and evolution facilitated by surrounding gas \citep{yangHierarchicalBlackHole2019,mckernanMonteCarloSimulations2020}, this environment significantly enhances the formation of massive BHs \citep{samsingAGNPotentialFactories2022}. Furthermore, recent modeling efforts suggest that compact binaries formed in AGN accretion disks can retain measurable eccentricities within the LIGO-Virgo-KAGRA (LVK) observational frequency band, making eccentricity detection a crucial diagnostic for verifying disc-mediated binary formation mechanisms \citep{samsingAGNPotentialFactories2022,fabjEccentricMergersAGN2024}. Intriguingly, the rotational symmetry of accretion disks has been proposed as a plausible explanation for the observed anti-correlation between the effective spin parameter $\chi_{\mathrm{eff}}$ and mass ratio $q$ in the LVK merger population \citep{2021ApJ...922L...5C,santiniBlackholeMergersDisklike2023,liResolving$h_rmEff$$q$2025}, suggesting that disk environments may imprint specific correlations on binary parameters.

Unlike mergers in other channels, BBH mergers in AGN accretion disks offer a unique opportunity to detect electromagnetic (EM) counterparts \citep{bartosRapidBrightStellarmass2017}. The mechanisms responsible for generating EM signals typically involve the interactions between the binary (or its merger remnant) and the surrounding medium, usually the gas in the AGN disks, such as ram-pressure stripping of gas around the remnant \citep{mckernanRampressureStrippingKicked2019}, hyper-Eddington accretion leading to a Bondi explosion \citep{wangAccretionmodifiedStarsAccretion2021} and jet launching after the post-merger kick \citep{chenElectromagneticCounterpartsPowered2024}. These processes result in delayed electromagnetic radiation that is superimposed on the AGN's intrinsic emission, making it detectable in observations.

Detecting EM counterparts to BBH mergers in AGN accretion disks holds transformative potential for multi-messenger astrophysics. These counterparts provide critical insights into both the merger dynamic and the ambient AGN environment \citep{vajpeyiMeasuringPropertiesActive2022}. Furthermore, secure association between GW events with specific AGN at confirmed redshift transforms these events into standard sirens, which enable independent constraints on the Hubble constant $H_{0}$ \citep{chenStandardSirenCosmological2022,mukherjeeFirstMeasurementHubble2020}. A statistically significant sample of such events could resolve the current Hubble tension \citep{divalentinoRealmHubbleTension2021} by providing an independent pathway to measure $H_{0}$.

Motivated by these scientific prospects, dedicated searches for such EM counterparts have been conducted through wide-field surveys. The Zwicky Transient Facility \citep[ZTF,][]{bellmZwickyTransientFacility2018,grahamZwickyTransientFacility2019} has identified seven unusual AGN flares that are temporally and spatially coincident with GW events observed during the O3 run, potentially representing EM counterparts to BBH mergers \citep{grahamLightDarkSearching2023}. Among these, the most promising candidate is event ZTF19abanrhr \citep{grahamCandidateElectromagneticCounterpart2020}, which occurred in AGN J124942.30+344928.9 approximately 30 days after GW190521, a BBH merger notable for its exceptionally high mass \citep{abbottGW190521BinaryBlack2020}.

However, these associations remain contentious due to observational limitations. Delayed spectroscopic follow-up observations failed to capture asymmetric broad-line features, which are predicted to arise when off-center flares unevenly illuminate the broad-line region \citep{mckernanRampressureStrippingKicked2019,cabreraSearchingElectromagneticEmission2024a}. Additionally, the original $\sim$ 3-year temporal baseline (2018-2021) may be insufficient to distinguish transient merger signals from stochastic AGN variability. The newly available ZTF photometry (2021-2024) now enables a statistically robust re-evaluation. Extended 6-year light curves allow us to examine AGN variability over a long timescale, thereby providing a more precise assessment of the nature of these flares. Furthermore, hierarchical merger models predict that a BBH remnant kicked out of the disk may return after roughly half an orbital period, triggering recurrent flares after $O(\mathrm{years})$ \citep{grahamCandidateElectromagneticCounterpart2020}, which we could search for in the extended data set.

If a flare is confirmed to be a genuine flare rather than AGN intrinsic variability, the next question is whether its temporal and spatial coincidence with a GW event reflects a physical association or chance coincidence. For the pair of ZTF19abanrhr and GW190521, Bayesian analyses yield conflicting conclusions: \citet{ashtonCurrentObservationsAre2021} found insufficient evidence for association mainly due to the AGN's position near the edge of the discovery skymap, whereas \citet{mortonGW190521BinaryBlack2023a} demonstrated strong evidence using updated skymap and hierarchical merger priors. For the remaining pairs of AGN and GW event, however, no systematic association studies exist, leaving their potential connections largely unexplored.

In this work, we expand the Bayesian framework presented in \citet{mortonGW190521BinaryBlack2023a} to quantify the probability of association for each pair of AGN flares and GW events proposed by \citet{grahamLightDarkSearching2023}. The remaining part of this paper is organized as follows. In Section \ref{sec:flares}, we present the optical light curves used in our analysis and calculate the significance of the flares. In Section \ref{sec:data}, we describe the GW data used in our analysis and calculate the kick velocities of the GW events. In Section \ref{sec:method}, we outline the theoretical framework used to evaluate the potential associations between AGN flares and GW events, and the results for each candidate pair are presented in Section \ref{sec:results}. In Section \ref{sec:hubble}, we estimate the Hubble constant with several possible pairs. Finally, we discuss the implications of our results and draw conclusions in Section \ref{sec:discussion}.

\section{AGN flares during O3 \label{sec:flares}}
\subsection{ZTF light curves and data processing}
ZTF is a time-domain survey with a 47 $\mathrm{deg}^{2}$ field-of-view camera mounted on the Palomar 48-inch Samuel Oschin Schmidt telescope \citep{bellmZwickyTransientFacility2018}, providing critical insights into transient phenomena. In \citet{grahamLightDarkSearching2023}, ZTF Data Release 5 was utilized to identify seven AGNs exhibiting flaring behavior that could be linked to BBH mergers. Our analysis employs the ZTF DR23 dataset, released on 2025 January 21, which includes all data from public surveys up to 2024 October 31 and all data from private surveys before 2023 June 30. This represents a three-year temporal extension compared to DR5, substantially improving our ability to distinguish genuine AGN flares from stochastic variability through extended baseline monitoring. ZTF surveys the entire observable northern sky every 2-3 days in the g- and r-bands, reaching a limiting magnitude of approximately 20.5 \citep{masciZwickyTransientFacility2018, bellmZwickyTransientFacility2019}, which ensures both sufficient sampling and sensitivity to detect transient features in our target AGNs.

\begin{figure}[htb!]
  \center
  \gridline{
    \includegraphics[trim=30 30 30 50, clip, width=0.5\textwidth]{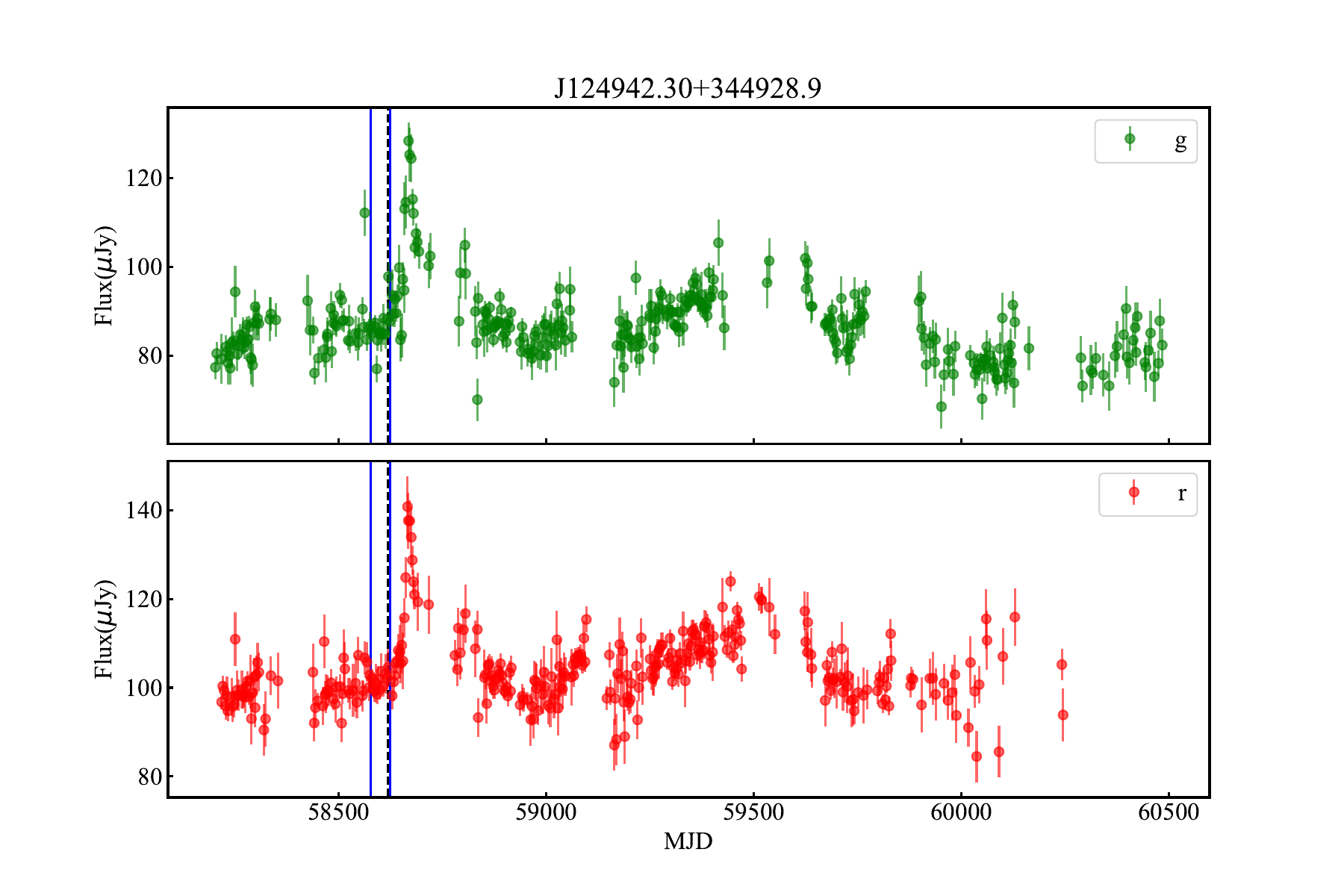}
    \includegraphics[trim=30 30 30 50, clip, width=0.5\textwidth]{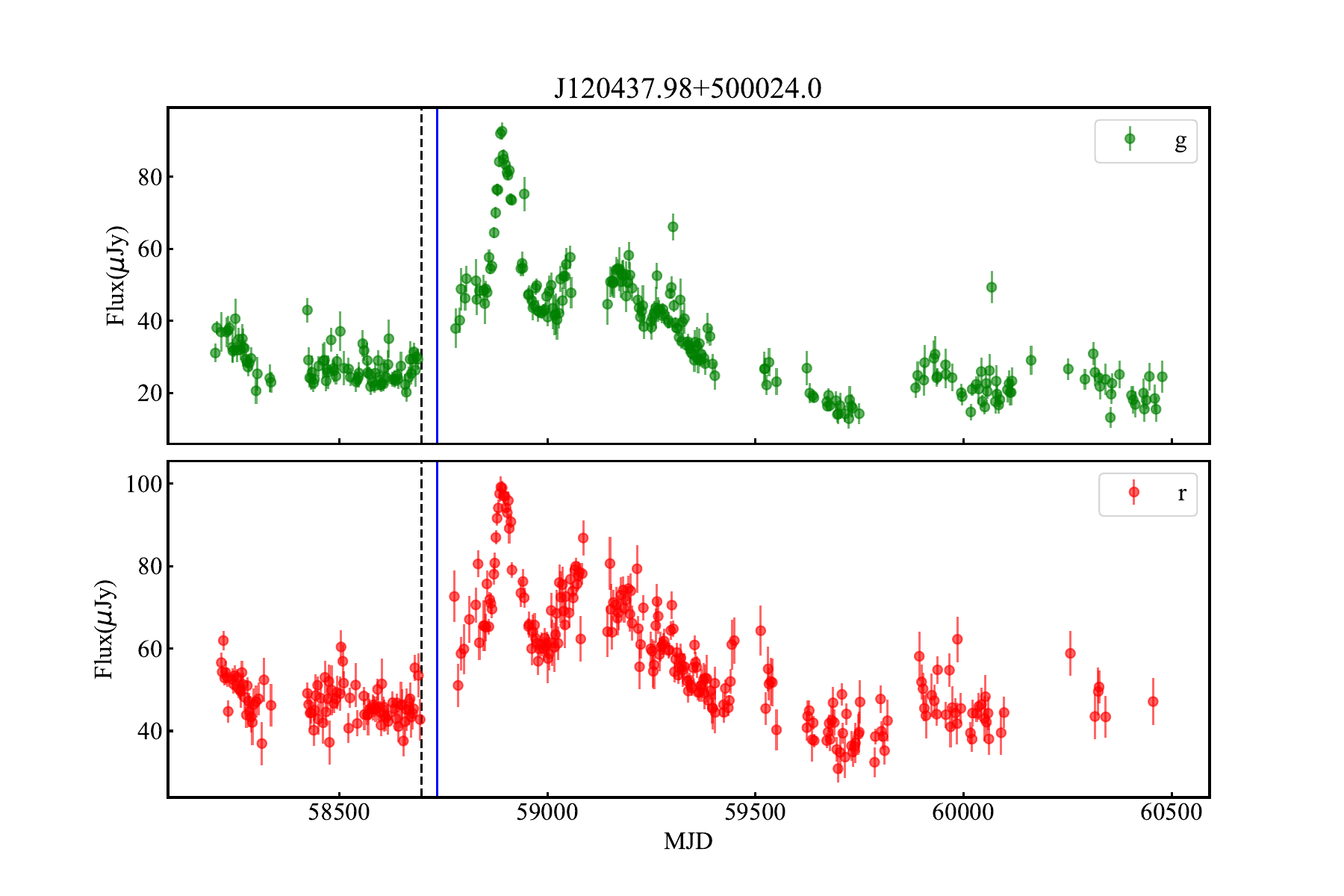}
  }
  \gridline{
    \includegraphics[trim=30 30 30 50, clip, width=0.5\textwidth]{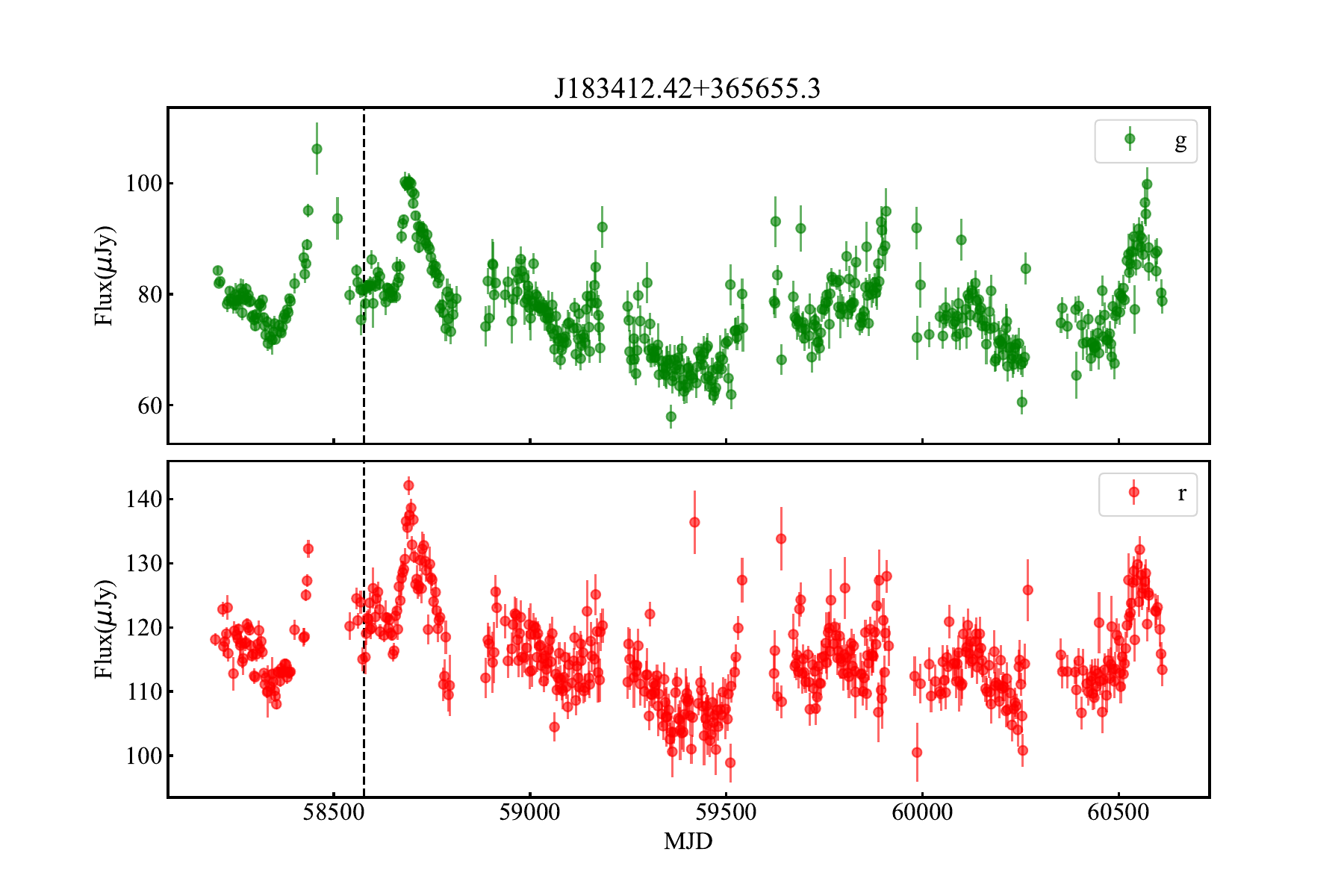}
    \includegraphics[trim=30 30 30 50, clip, width=0.5\textwidth]{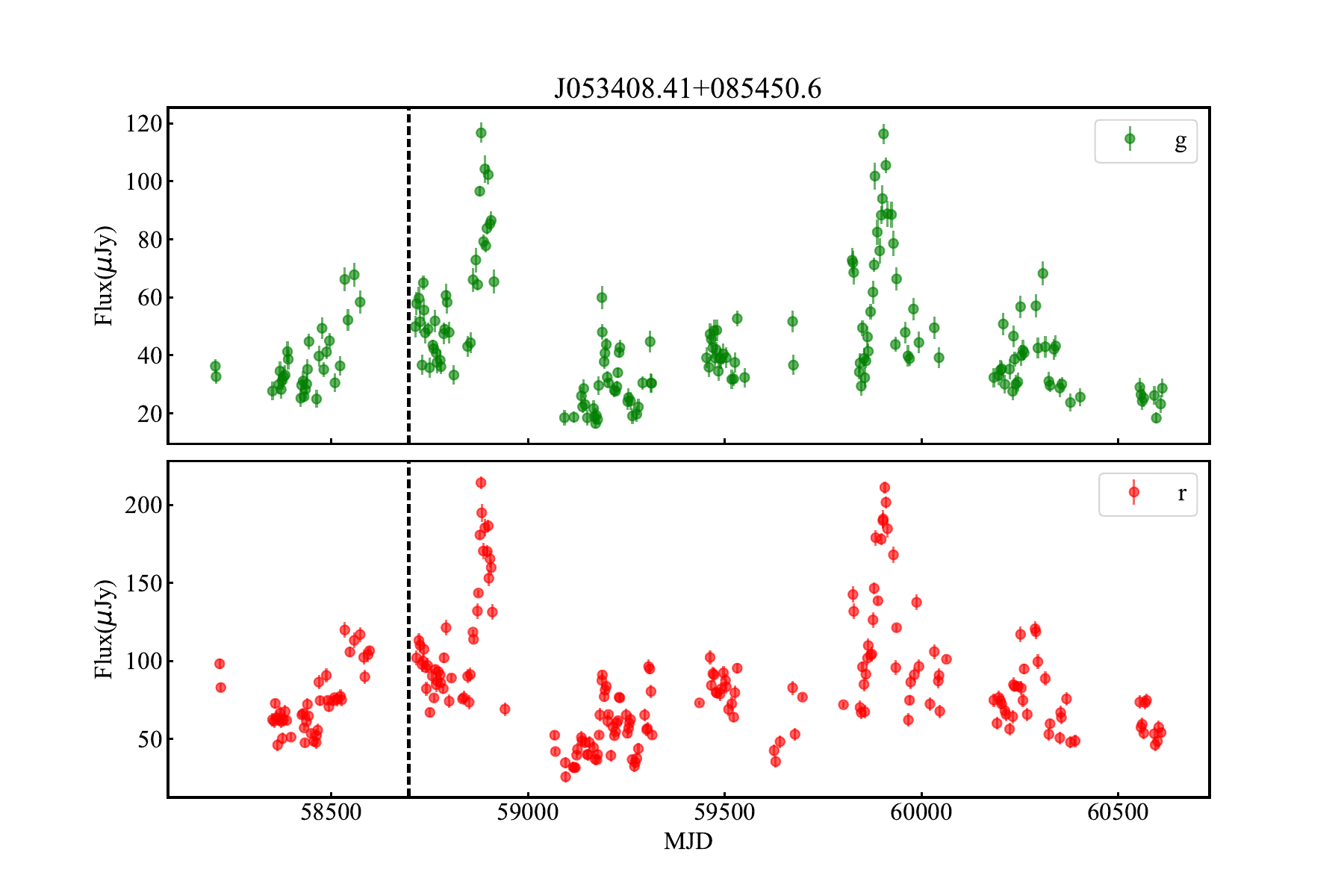}
  }
  \gridline{
    \includegraphics[trim=30 30 30 50, clip, width=0.5\textwidth]{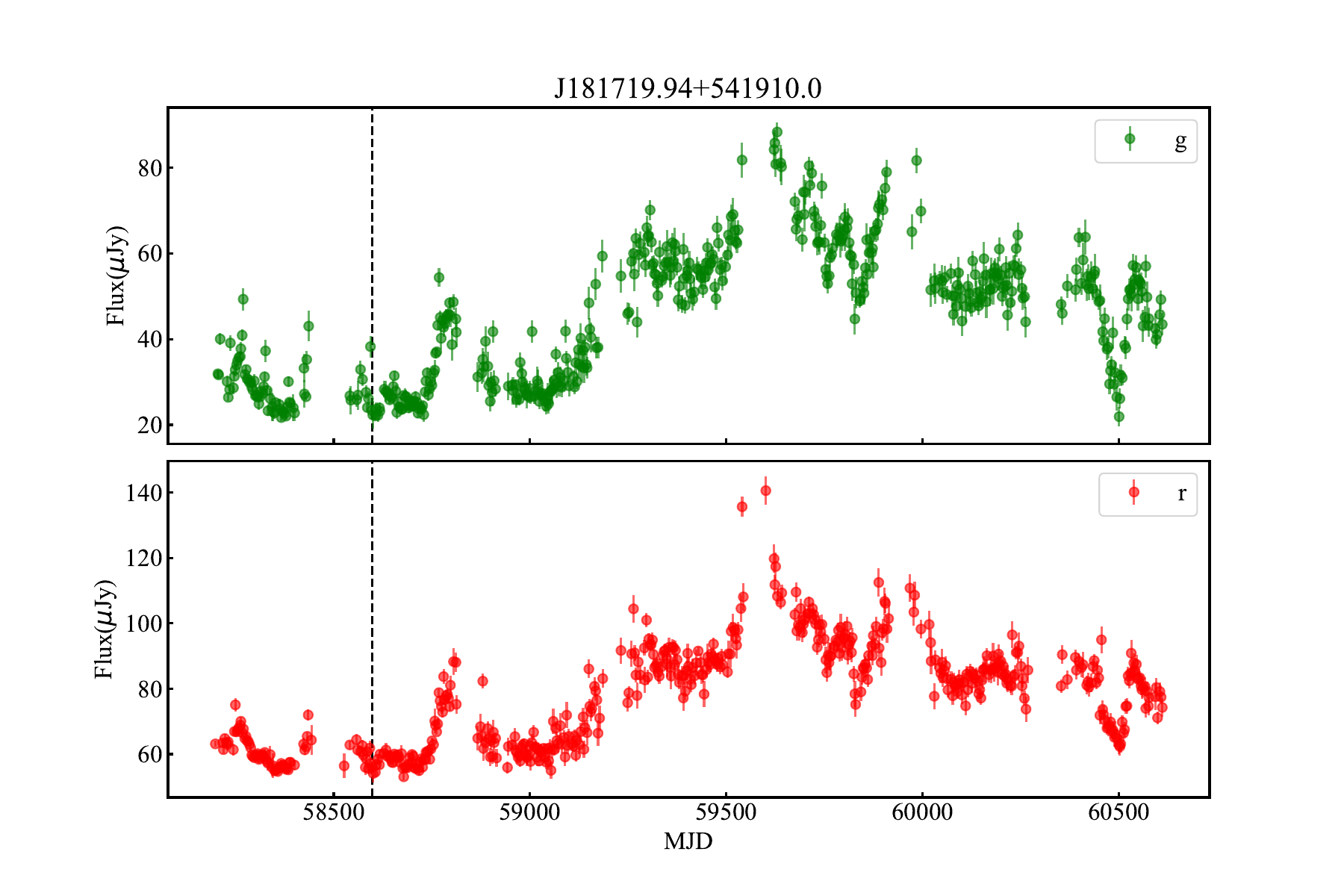}
    \includegraphics[trim=30 30 30 50, clip, width=0.5\textwidth]{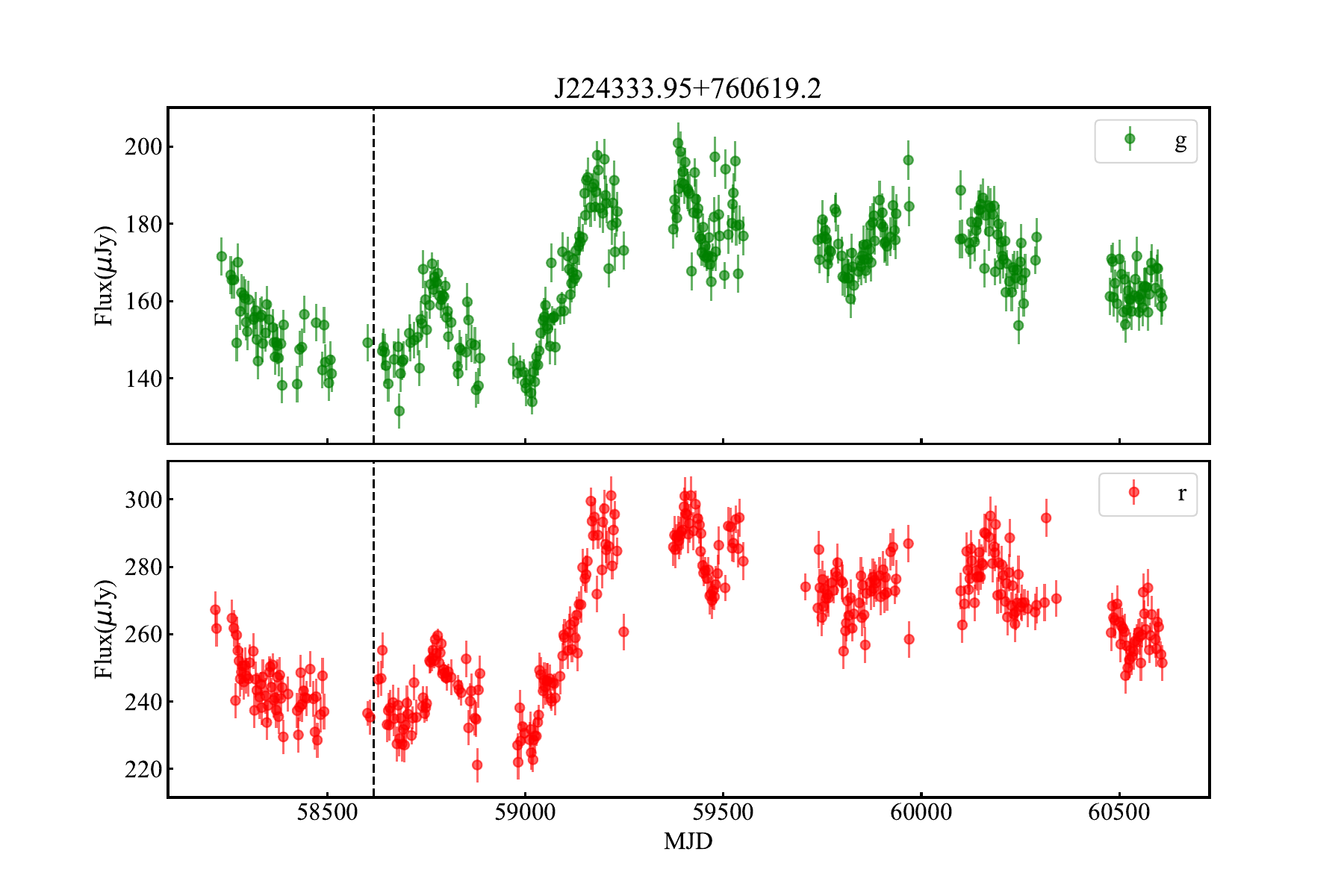}
  }
  \includegraphics[trim=20 30 30 50, clip, width=0.5\textwidth]{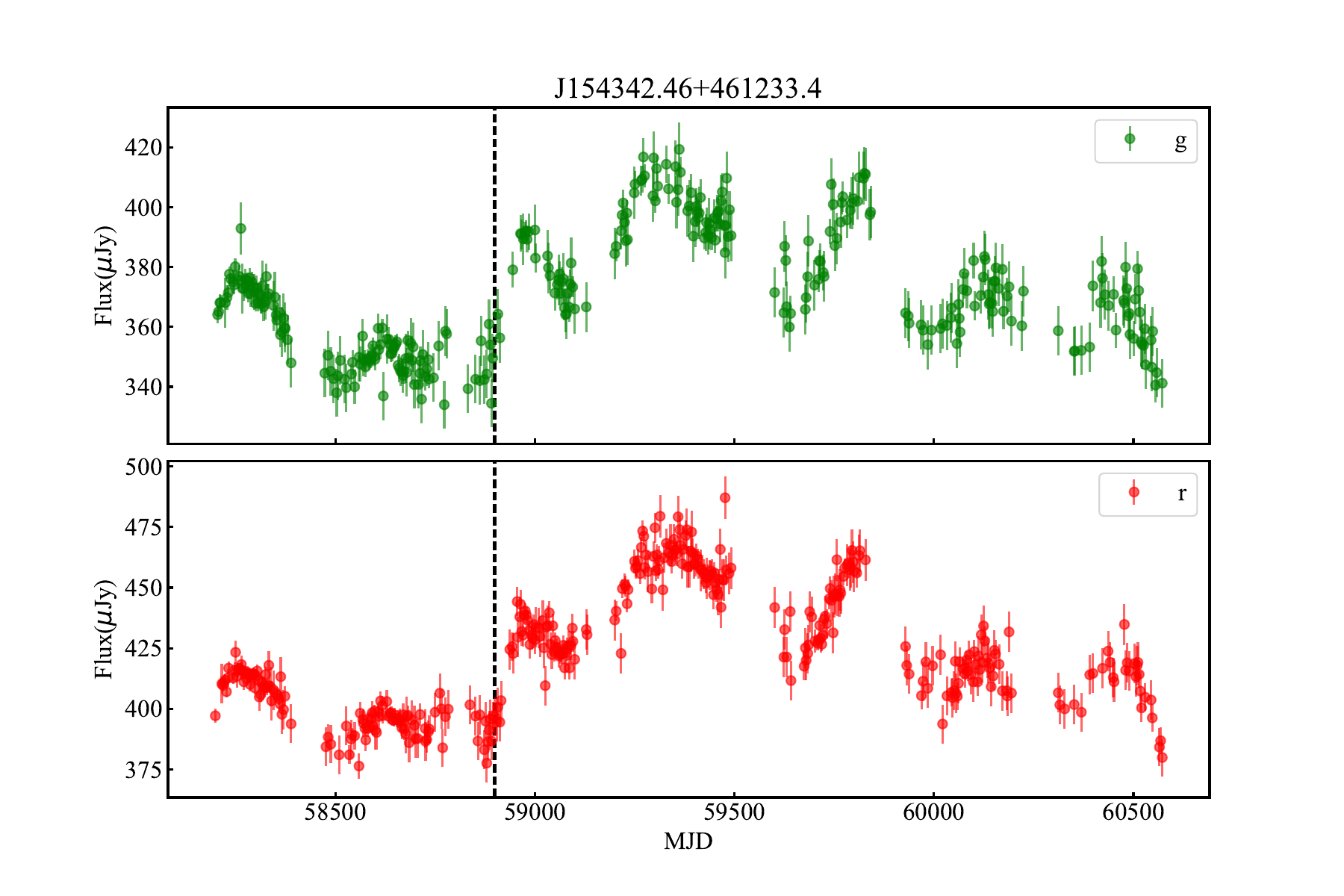}
  \caption{The ZTF g- and r-band light curves for the seven AGNs associated with GW events. The vertical lines represent the trigger times of the GW events. The blue lines indicate the events that remain as possible associations after our analysis, while associations with all other events have been excluded.  }
  \label{fig:lightcurves}
\end{figure}

To mitigate contamination while preserving astrophysical signals, we implement a hybrid sigma-clipping algorithm. A standard 3$\sigma$ clipping is applied first, which removes outliers that deviate $> 3\sigma$ from the median flux. Then, to ensure that the potential flare signals are retained, we preserve sequences where more than two consecutive points exceed 3$\sigma$ and isolated $3\sigma$ outliers flanked by points $>2.5\sigma$ (characteristic of flare peaks). After cleaning, the light curves are binned into 3-day intervals to improve the signal-to-noise ratio (SNR). The final light curves for the seven AGNs are shown in Figure \ref{fig:lightcurves}.

\subsection{AGN flare significance analysis via Gaussian Processes \label{sec:pflare}}

To quantify the statistical significance of candidate flares against the stochastic variability inherent in AGNs, we introduce the metric $p_{\mathrm{flare}}$, defined as the probability that a given flare is indeed a genuine flare rather than a characteristic of variability intrinsic to the AGN itself.

We calculate $p_{\mathrm{flare}}$ with Gaussian Processes (GPs), a class of non-parametric statistical models widely used for regression and classification tasks, particularly when dealing with complex and noisy data. Formally, a GP defines a distribution over functions, where any finite collection of function values follows a joint multivariate Gaussian distribution. GPs have become valuable tools in astronomy, particularly for modeling and analyzing time series data with inherent stochasticity \citep{aigrainGaussianProcessRegression2023}, such as quasi-periodic oscillations \citep{yangGaussianProcessModeling2021}, exoplanet transits \citep{crossfield197CANDIDATES1042016} and AGN variability \citep{kozlowskiQUANTIFYINGQUASARVARIABILITY2009,macleodMODELINGTIMEVARIABILITY2010}.

A GP is completely specified by its mean function and covariance function. For the mean function, we use the median of the light curve, which is sufficiently effective in most AGN cases. For the covariance function, commonly known as the kernel, we select the Matern-1/2 kernel, which has proven to be effective in modeling AGN light curves \citep{griffithsModelingMultiwavelengthVariability2021}. The kernel is defined as:
\begin{equation}
k(t,t') = \rho^{2}\exp\left(-\frac{|t-t'|}{\tau}\right),
\end{equation}
where $\rho$ is the variability amplitude hyperparameter, and $\tau$ is the variability timescale hyperparameter. Given $N$ observations of a variable $\boldsymbol{y} = \{y_{i}\}_{i=1,...,N}$ with associated measurement uncertainties $\boldsymbol{\sigma} = \{\sigma_{i}\}$, which are taken at times $\boldsymbol{t}=\{t_{i}\}$, the marginal log likelihood is given by:
\begin{equation}
  \log \mathcal{L} = -\frac12(\boldsymbol{y}-\boldsymbol{m})^{T}\boldsymbol{K}^{-1}(\boldsymbol{y}-\boldsymbol{m})-\frac12 \log |\boldsymbol{K}| - \frac N2\log 2\pi,
\end{equation}
where the mean vector $\boldsymbol{m}$ consists of identical elements, each equal to the median of the data. The term $\boldsymbol{K}$ is a matrix with elements $K_{ij} = k(t_{i},t_{j}) +\delta_{ij}\sigma_{i}^{2}$, where $\delta_{ij}$ is the discrete Kronecker delta function. For a given light curve, the optimal hyperparameters are obtained by maximizing the likelihood function, which gives the best-fit values that provide the most accurate description of the observed data under the GP model.

If a light curve contains a flare within a specific time interval $t_{L}$, where $t_{L} = \{t_{peak} - L/2 \le t \le t_{peak} + L/2\}$ with $t_{peak}$ being the flare peak time and $L$ interval length, the hyperparameters fitted to the non-flare segments, denoted as $\rho_{0}$ and $\tau_{0}$, cannot accurately capture the variability during the flare. This is because the GP model trained on the non-flare data is designed to describe the baseline variability and does not account for the distinct features introduced by the flare. A test statistic, similar to that used in \citet{grahamLightDarkSearching2023}, is defined as:
\begin{equation}
  \lambda = (\boldsymbol{y}_{L}-\boldsymbol{m}_{0})^{T}\boldsymbol{K}_{0}^{-1}(\boldsymbol{y}_{L}-\boldsymbol{m}_{0}),
\end{equation}
where $\boldsymbol{y}_{L}$ represents the observations at times $\boldsymbol{t}_{L}$ and $\boldsymbol{m}_{0}$ is determined from the non-flare data, matching the length of $\boldsymbol{y}_{L}$. In the analysis, we set $L$ to 50 days, consistent with the approach in \citet{grahamLightDarkSearching2023}.

For each AGN light curve with a potential flare, we first estimate the hyperparameters $\tau_{0}$ and $\rho_{0}$ using the non-flare data. Using these hyperparameters, we generate 10,000 simulated light curves with GPs, matching the observation time and uncertainties. Subsequently, we compute the test statistic $\lambda$ for both the observed light curve and the simulated light curves. By comparing the observed $\lambda$ with the distribution of $\lambda$ values from the simulations, we obtain the $p_{\mathrm{flare}}$ for each AGN.

\begin{figure}[htb]
   \gridline{
    \includegraphics[trim=30 10 30 50, clip, width=0.5\textwidth]{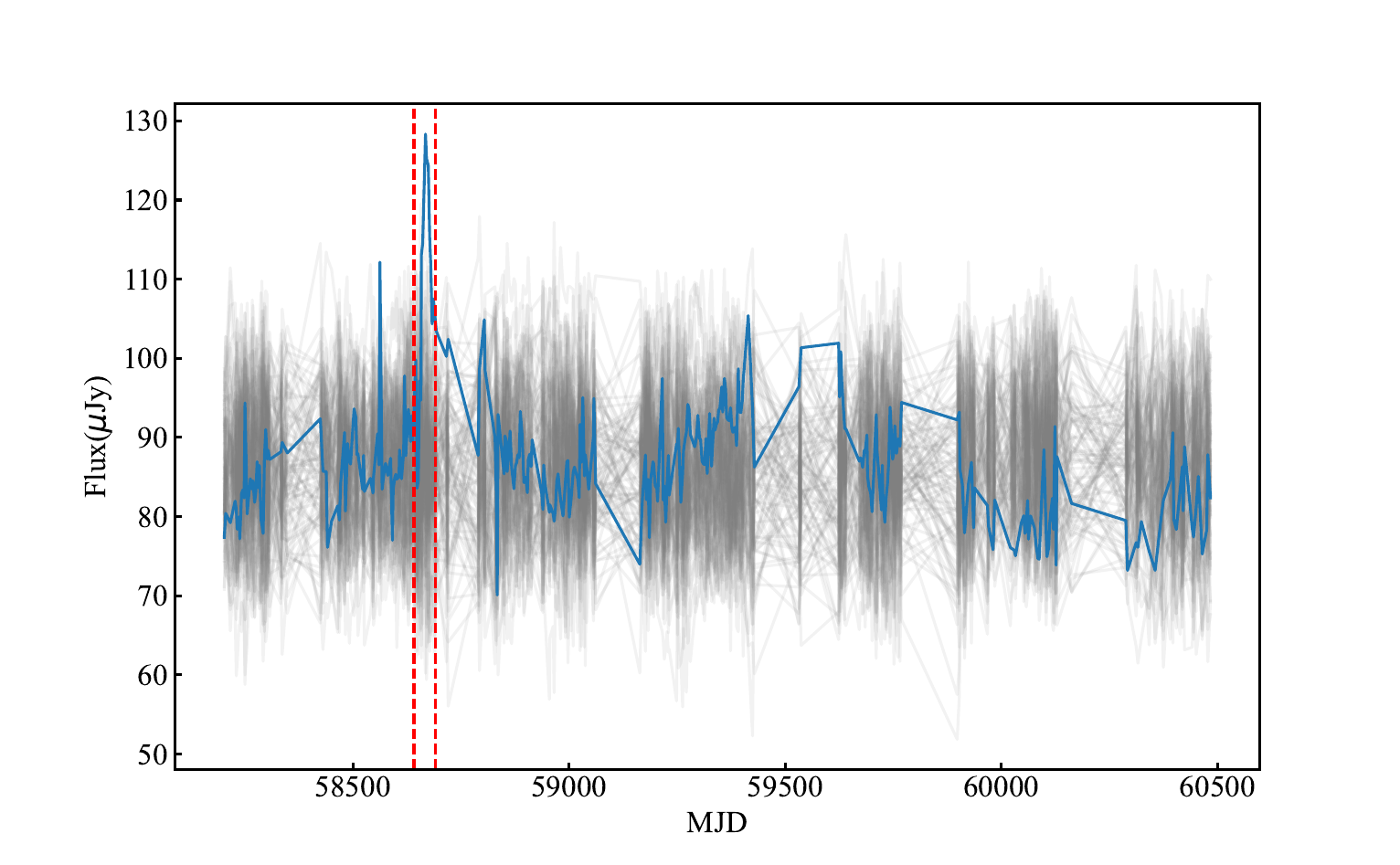}
    \includegraphics[trim=30 10 30 50, clip, width=0.5\textwidth]{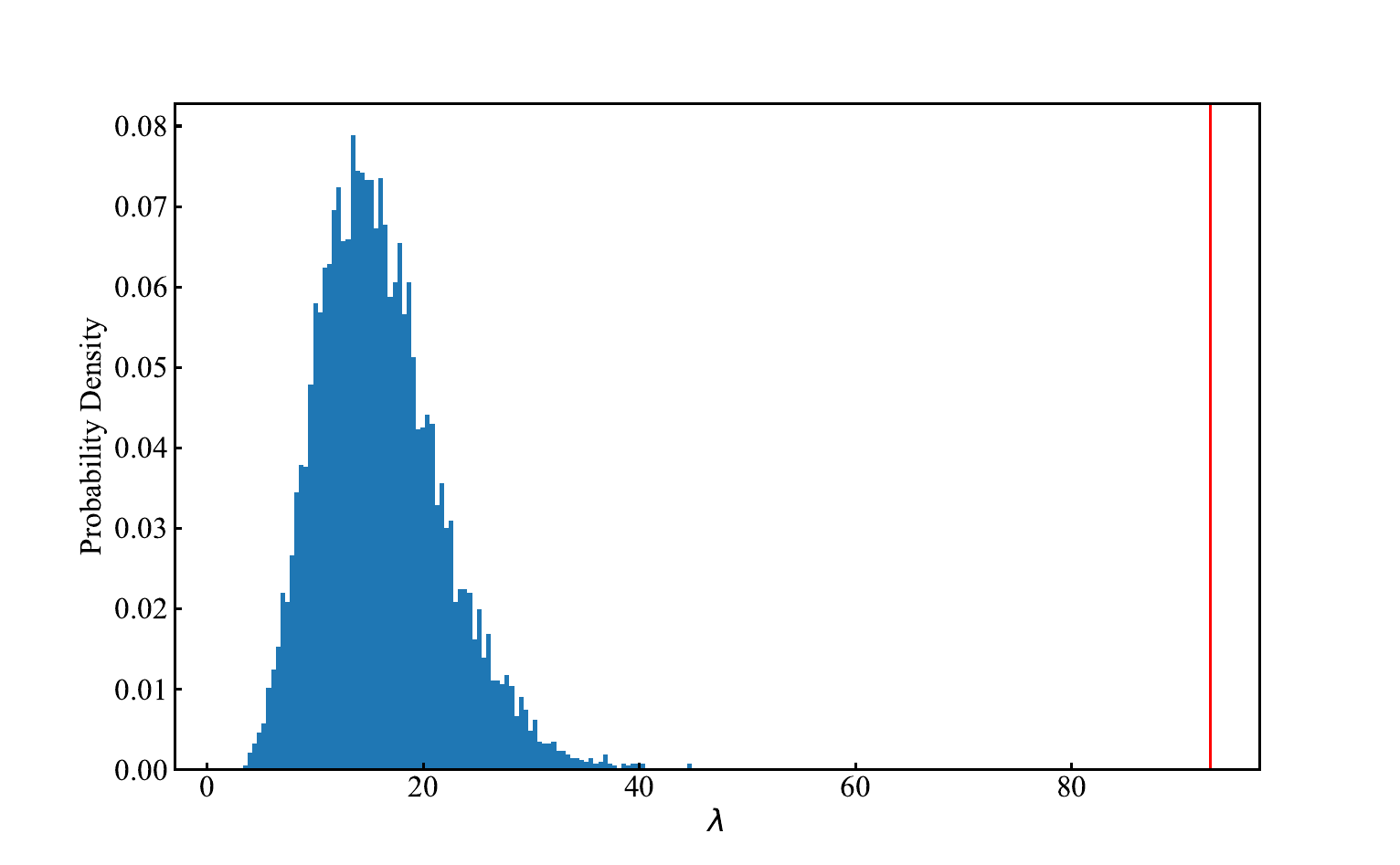}
  }
  \caption{(left) A real AGN light curve (blue line) and 100 random light curves (gray lines) generated with the same GP model fitted to the AGN data. The two red vertical dashed lines indicate the selected time interval for computing the $\lambda$ value. (right) The distribution of $\lambda$ values from the simulated light curves, with the red vertical line representing $\lambda$ value from the real AGN.}
  \label{fig:GP}
\end{figure}

The above steps are implemented with \texttt{celerite} \citep{foreman-mackeyFastScalableGaussian2017}, a Python package designed for efficient and scalable Gaussian Process modeling and analysis. Figure \ref{fig:GP} shows a real AGN light curve along with several light curves generated using the same GP model fitted to the AGN. In addition, the $\lambda$ values for each light curve are computed. The $\lambda$ value of the real AGN significantly deviates from the distribution of $\lambda$ values obtained from the simulated light curves, indicating the presence of a flare that is not attributable to typical AGN variability.

For each AGN, $p_{\mathrm{flare}}$ is independently calculated using the g- and r-band light curves, with the smaller value adopted to conservatively minimize false positives (Table \ref{table:GP}). Our analysis identifies three AGNs (J181719.94+541910.0, J224333.95+760619.2 and J154342.46+461233.4) with $p_{\mathrm{flare}} < 0.995$, falling below the empirical threshold used in \citet{grahamLightDarkSearching2023}. This downward revision in significance arises
from the extended temporal baseline of ZTF DR23 (spanning three additional three years compared to DR5), which resolves previously flagged flares as natural extremes of AGN stochastic variability. Of the remaining four candidates, the AGN J053408.41+085450.6 has a radio counterpart at 21 cm in the NRAO VLA Sky Survey \citep{condonNRAOVLASky1998}, suggesting a blazar classification. Given that such flares are common in blazars \citep{guOpticalVariabilityFlatspectrum2011}, we do not consider this AGN to be associated with any GW events. Therefore, only three AGN flares remain as possible counterparts for further analysis.

\begin{deluxetable}{ccccccc}[htb]
  \tablecaption{Summary of the 7 AGN flares\label{table:GP}}
  \tablehead{
    \colhead{AGN Name} & \colhead{Redshift} & \colhead{$\log_{10}(M_{\mathrm{SMBH}})$} & \colhead{$t_{\mathrm{peak}}$} & \colhead{$p_{\mathrm{g}}$} & \colhead{$p_{\mathrm{r}}$} & \colhead{$p_{\mathrm{flare}}$} \\
   & &\colhead{($M_{\odot}$)} & \colhead{(MJD)}}
  \startdata
  J124942.30+344928.9 & 0.438 & 8.6 & 58666 & 0.9999 & 0.9999 & 0.9999\\
  J120437.98+500024.0 & 0.389 & 8.0 & 58890 & 0.9999 & 0.9999 & 0.9999\\
  J183412.42+365655.3 & 0.419 & 9.1 & 58690 & 0.9994 & 0.9994 & 0.9994\\
    \multicolumn{7}{c}{\hdashrule[0.5ex]{13cm}{1pt}{3pt}} \\
  J053408.41+085450.6 & 0.5 & (8.0) & 58880 & 0.9999 & 0.9998 & 0.9998\\
  J181719.94+541910.0 & 0.234 & 8.0 & 58805 & 0.2521 & 0.9592 & 0.2521\\
  J224333.95+760619.2 & 0.353 & 8.8 & 58775 & 0.2442 & 0.6265 & 0.2442\\
  J154342.46+461233.4 & 0.599 & 9.3 & 58960 & 0.0902 & 0.4251 & 0.0902\\
  \enddata
  \tablecomments{The redshift and the SMBH mass ($M_{\mathrm{SMBH}}$) are taken from \citet{grahamLightDarkSearching2023}, with the redshift for J053408.41+085450.6 being a photometric redshift and its SMBH mass assumed to be the fiducial value of $10^{8} M_{\odot}$. $t_{\mathrm{peak}}$ is the peak time of the flare. $p_{\mathrm{g}}$ and $p_{\mathrm{r}}$ represent the probabilities that the AGN flare exists in the g or r band, respectively, and $p_{\mathrm{flare}}$ is the smaller one of the two.
  Events above the horizontal dashed line are those identified as AGN flares in this work.}
\end{deluxetable}

\section{Gravitational Wave Data \label{sec:data}}
\subsection{GW posterior probability density}

The LVK Collaboration released the Gravitational Wave Transient Catalog (GWTC), which includes posterior distribution samples\footnote{ https://gwosc.org} for the parameters of each GW event. The O3a and O3b events are listed in GWTC-2.1 \citep{abbottGWTC21DeepExtended2024} and GWTC-3 \citep{abbottGWTC3CompactBinary2023}, respectively, with their \textsc{Mixed} posterior samples used in this work. However, two events, GW190424\_180648 and GW190909\_114149, fall below the threshold in GWTC-2.1. For these events, we use the publication samples from GWTC-2 \citep{abbottGWTC2CompactBinary2021}. Several key parameters for the GW events used in this analysis are given in Table \ref{table:GW}.

\begin{deluxetable}{cccccccccc}[htb]
  \tablecaption{Summary of the parameters for the 9 GW events\label{table:GW}}
  \tablehead{
    \colhead{GW ID} & \colhead{Data set} & \colhead{$D_{\mathrm{L}}$} &\colhead{ $M_{1}$ } & \colhead{$M_{2}$} & \colhead{$\chi_{1}$} & \colhead{$\chi_{2}$} & \colhead{$v_{\mathrm{k}}$} &\colhead{$A_{90}$} & \colhead{$n_{s}$}\\
    & & \colhead{(Gpc)} & \colhead{($M_{\odot}$)} & \colhead{($M_{\odot}$)} &
    \colhead{} &
    \colhead{}&
    \colhead{($\mathrm{km \ s^{-1}}$)}  & \colhead{($\deg^{2}$)} }
  \startdata
  GW190403\_051519 & GWTC-2.1 & $8.3^{+6.7}_{-4. 3}$ & $85^{+28}_{-33}$ & $20^{+26}_{-8}$  & $0.89^{+0.09}_{-0.31}$ & $0.53^{+0.43}_{-0.47}$ & $510^{+1300}_{-420}$ & 3930 & 94466\\
  GW190521 & GWTC-2.1 & $3.3^{+2.8}_{-1.8} $ & $98^{+34}_{-22}$ & $57^{+27}_{-30}$ & $0.71^{+0.26}_{-0.63}$ & $0.53^{+0.42}_{-0.48}$ & $730^{+1400}_{-570}$ & 1021 & 16221\\
  GW190803\_022701 & GWTC-2.1 & $3.2^{+1.6}_{-1.5}$ & $38^{+10}_{-7}$ & $28^{+8}_{-8}$ & $0.41^{+0.50}_{-0.38}$ & $0.44^{+0.48}_{-0.40}$ & $480^{+1100}_{-380}$ & 1004 & 7315\\
  \multicolumn{10}{c}{\hdashrule[0.5ex]{16cm}{1pt}{3pt}} \\
  GW190424\_180648 & GWTC-2 & $2.6^{+1.6}_{-1.3}$ & $39^{+11}_{-7}$ & $31^{+7}_{-7}$ & $0.54^{+0.41}_{-0.48}$ & $0.47^{+0.47}_{-0.42}$ & $640^{+1300}_{-540}$ & 28336 & 86909\\
  GW190514\_065416 & GWTC-2.1 & $3.9^{+2.6}_{-2.1}$ & $41^{+17}_{-9}$ & $28^{+10}_{-10}$ & $0.47^{+0.47}_{-0.42}$ & $0.48^{+0.47}_{-0.43}$ & $470^{+1000}_{-350}$ & 3435 & 30861\\
  GW190731\_140936 & GWTC-2.1 & $3.3^{+2.4}_{-1.8}$ & $42^{+13}_{-9}$ & $29^{+10}_{-10}$ & $0.39^{+0.52}_{-0.35}$ & $0.46^{+0.47}_{-0.42}$ & $500^{+1300}_{-390}$  & 3640 & 16688\\
  GW190909\_114149 & GWTC-2 & $4.9^{+3.7}_{-2.6}$ & $43^{+51}_{-12}$ & $28^{+13}_{-11}$ & $0.60^{+0.36}_{-0.54}$ & $0.49^{+0.46}_{-0.44}$ & $470^{+1100}_{-350}$ & 4732 & 42047\\
  GW200216\_220804 & GWTC-3 & $3.8^{+3.0}_{-2.0}$ & $51^{+22}_{-13}$ & $30^{+14}_{-16}$ & $0.48^{+0.46}_{-0.43}$ & $0.51^{+0.44}_{-0.46}$ & $490^{+1300}_{-370}$ & 3010 & 34563\\
  GW200220\_124850 & GWTC-3 & $4.0^{+2.8}_{-2.2}$ & $39^{+14}_{-9}$ & $28^{+9}_{-9}$ & $0.51^{+0.43}_{-0.46}$ & $0.47^{+0.46}_{-0.43}$ & $540^{+1100}_{-420}$ & 3169 & 33500
  \enddata
  \tablecomments{ Luminosity distance ($D_{\mathrm{L}}$), component masses ($M_{1,2}$) and component spins ($\chi_{1,2}$) are obtained from GWTC. The velocity of the kick ($v_{\mathrm{k}}$) is inferred using the method described in \citet{varmaExtractingGravitationalRecoil2020}. The uncertainties correspond to the 90\% credible interval. $A_{90}$ is the 90\% confidence area of the event, and  $n_{s}$ is the number of Million Quasar Catalog \citep[v8,][]{fleschMillionQuasarsMilliquas2023} sources within the 90\% confidence volume. Events above the horizontal dashed line are those identified in this work that may potentially have EM counterparts.  }
\end{deluxetable}

These posterior samples are used to derive posterior probability distributions with a Dirichlet process Gaussian mixture model (DPGMM), which is a non-parametric Bayesian method that allows flexible approximation of the probability distributions \citep{nguyenApproximationFiniteMixtures2020}. To obtain posterior distributions, we make use of the open-source Python package \texttt{FIGARO}\footnote{https://github.com/sterinaldi/FIGARO} \citep{rinaldiRapidLocalizationGravitational2022}, which employs a DPGMM to make an accurate approximation of the distributions from the available samples.

\subsection{Kick velocity analysis}

During the final stages of BBH coalescence, GW emission impacts a recoil or kick to the remnant BH, predominantly near the merger phase \citep{gonzalezMaximumKickNonspinning2007}. This kick plays a critical role in determining whether the remnant remains gravitational bound to its host environment or escapes, influencing hierarchical merger rates and electromagnetic counterpart detectability.

Following the method in \citet{varmaExtractingGravitationalRecoil2020}, we compute the kick velocity ($v_{\mathrm{k}}$) using the surrogate models \texttt{NRSur7dq4} and \texttt{NRSur7dq4Remnant} \citep{varmaSurrogateModelsPrecessing2019}. The posterior samples of binary properties including component masses and 3D spin vectors are extracted from the GWTC for events whose parameter estimation results using the waveform \texttt{NRSur7dq4} are contained in the release. Otherwise, we rerun parameter estimation with this waveform model. Then the kick velocity can be evaluated using the \texttt{surfinBH}\footnote{https://doi.org/10.5281/zenodo.1435832} \added{\citep{https://doi.org/10.5281/zenodo.1435832}}.

The inferred $v_{\mathrm{k}}$ values for all GW events are summarized in Table \ref{table:GW}, showing consistency with the upper and lower limits reported in \citet{grahamLightDarkSearching2023}. Crucially, these values remain orders of magnitude below the Keplerian orbital velocity of BBHs in the AGN disks ($v_{\mathrm{orb}}\sim10^{4}(r/10^{3}R_{\mathrm{s}})^{-1/2}\ \mathrm{ km \ s^{-1}}$, $R_{s}$ is the Schwarzschild radius of the central supermassive black hole), indicating that this kick has no significant impact on the orbital and is insufficient to cause the remnant to escape the AGN. Therefore, repeat flares are expected for these GW events if they occur in the AGN disks.

\section{Association with GW events\label{sec:method}}
In \citet{grahamLightDarkSearching2023}, AGN flares are preliminarily flagged as potential GW counterparts if they locate within the 90\% confidence volume of a GW event and exhibit a peak occurring within 200 days post-GW trigger. Furthermore, a p-value approach is applied to estimate the chance coincidences given the observed matched pairs. However, this approach does not statistically evaluate individual associations.

To rigorously assess whether a flare and GW event originate from the same source and to identify the most probable GW event associated with a given flare, we employ a Bayesian statistical framework proposed by \citet{mortonGW190521BinaryBlack2023a}. This framework is used to assess whether the flare and a potential GW event share a common origin or are merely coincidental. The primary objective is to calculate the odds ratio $\mathcal{O}^{A}_{C}$ between two competing models:
\begin{equation}
 \mathcal{O}^{A}_{C} = \frac{p(\mathcal{H}_{A}|d)}{p(\mathcal{H}_{C}|d)}= \frac{p(d|\mathcal{H}_{A})}{p(d|\mathcal{H}_{C})}\frac{p(\mathcal{H}_{A})}{p(\mathcal{H}_{C})}=\mathcal{B}^{A}_{C}\mathcal{P}^{A}_{C}.
\end{equation}
In this equation, $d$ represents the data, while $\mathcal{H}_{A}$ and $\mathcal{H}_{c}$ denote the competing hypotheses: the association model (A) and the coincidence model (C), respectively. In the association model, the flare and the GW event share a common astrophysical origin, whereas in the coincidence model, the flare and the GW event are unrelated.

The prior odds $\mathcal{P}^{A}_{C}$ represents the initial belief or expectation before observing, based on theoretical information or previous knowledge. The Bayes factor $\mathcal{B}^{A}_{C}$ quantifies the support provided by the data for one model over the other, calculated as the ratio of their evidence. The odds ratio $\mathcal{O^{A}_{C}}$ combines the prior odds and the Bayes factor to provide a comprehensive assessment. A detailed explanation of these terms is provided in the following sections.

\subsection{Prior odds}

In most cases, the prior odds are set to unity, assuming equal likelihood for both models. This choice reflects a state of minimal prior knowledge or bias, ensuring that the evaluation of the models is driven solely by the evidence provided by the data. However, an approximate estimate of the previous odds has been found to be the inverse of the expected number $N$ of events that could plausibly be considered consistent with the association model within the specific duration and volume of search \citep{ashtonCoincidentDetectionSignificance2018}. Previous studies typically set $N$ to 13 \citep{ashtonCurrentObservationsAre2021}, which is the number of flares similar to ZTF19abanrhr in the ZTF alert stream \citep{grahamCandidateElectromagneticCounterpart2020}. However, it is important to note that using the same prior odds for every pair of GW and AGN is inappropriate, since the confidence regions of the GW localizations and the significance levels of the AGN flares can vary between events. In our analysis, $N$ is estimated as the number of AGNs within the localization region of the GW event that could potentially produce flares of similar significance \citep{ashtonCurrentObservationsAre2021}.

\begin{figure}[htb]
  \plotone{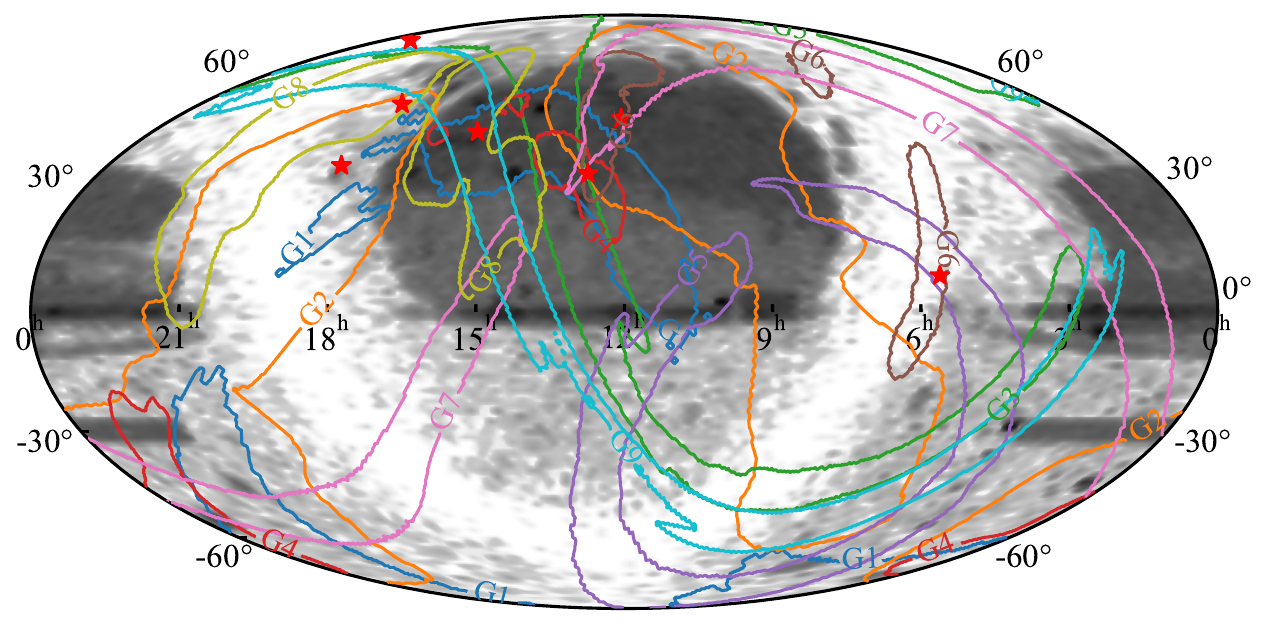}
  \caption{Position of 7 flares (red stars, from left to right: J183412+365655.3, J181719.94+541910.0, J224333.95+760619.2, J154342.46+461233.4, J124942.30+344928.9, J120437.98+500024.0 and J053408.41+085450.6) described in Section \ref{sec:flares} and 90 percent credibility level (CL) localization surfaces of the 9 possible associated GW events (solid lines): GW190403\_051519 (G1), GW190424\_180648 (G2), GW190514\_065416 (G3), GW190521 (G4), GW190731\_140936 (G5), GW190803\_022701 (G6), GW190909\_114149 (G7), GW200216\_220804 (G8) and GW200220\_124850 (G9). The background illustrates the spatial distribution of AGNs from the Million Quasars catalog.}
  \label{fig:AGN_GW}
\end{figure}

The prior odds are then calculated as:
\begin{equation}
  \mathcal{P}^{A}_{C} = \frac1N= \frac{1}{(1-p_{\mathrm{flare}})n_{\mathrm{s}}}.
\end{equation}
Here, $p_{\mathrm{flare}}$ is the probability that a given flare is indeed a genuine flare rather than a variability feature intrinsic to the AGN itself, which we calculated in Section \ref{sec:pflare}, and $n_{s}$ is the number of AGNs located within the 90\% confidence volume of the corresponding GW event. We estimate $n_{\mathrm{s}}$ with the Million Quasar Catalog \citep[v8,][]{fleschMillionQuasarsMilliquas2023}, which includes all published quasars to 2023 June 30. Following the procedure in \citet{grahamLightDarkSearching2023}, we exclude all sources at redshift $z>1.2$ and known blazars. The distribution of sources in this catalog is presented in Figure \ref{fig:AGN_GW}, as well as the position of flares and the 90\% confidence area of GW events used in this work. For each GW event, we calculate $n_{\mathrm{s}}$ with the \texttt{postprocess.crossmatch} function from the \texttt{ligo.skymap}\footnote{https://lscsoft.docs.ligo.org/ligo.skymap/} Python package, and the results are listed in Table \ref{table:GW}.

\subsection{Bayes factor}

The Bayes factor is defined as the ratio of the evidences:
\begin{equation}
  \mathcal{B}^{A}_{C}=\frac{p(d|\mathcal{H}_{A})}{p(d|\mathcal{H}_{c})}.
\end{equation}
We calculate $\mathcal{B}^{A}_{C}$ follows the formalism in \citet{mortonGW190521BinaryBlack2023a}, where the evidence for each model can be written as:
\begin{equation}
  p(d|\mathcal{H}_{i}) = \int \frac{p(M_{1}^{\mathrm{eff}},D_{L}^{\mathrm{eff}},\alpha,\delta | d)}{p(D_{L}^{\mathrm{eff}})} \times p(M_{1}^{\mathrm{eff}},D_{L}^{\mathrm{eff}},\alpha,\delta|\mathcal{H}_{i}) \mathrm{d}M_{1}^{\mathrm{eff}} \mathrm{d} D_{L}^{\mathrm{eff}} \mathrm{d}\alpha \mathrm{d}\delta,
\end{equation}
where $M_{1}^{\mathrm{eff}}$ is the detector-frame effective primary mass, $D_{L}^{\mathrm{eff}}$ is the effective luminosity distance and $\alpha$ and $\delta$ are the right ascension and declination of the transient source. Specifically, $p(M_{1}^{\mathrm{eff}},D_{L}^{\mathrm{eff}},\alpha,\delta |d)$ is the posterior probability density of the GW event and $p(D_{L}^{\mathrm{eff}})$, which follows \citet{farrSimpleLuminosityDistance2020}, is the prior probability corresponding to a uniform merger rate in the source frame for parameter estimation by the LVK collaboration.

In the association model $\mathcal{H}_{A}$, the GW event is assumed to originate from a specific AGN, so its position in the sky is fixed at ($\alpha_{\mathrm{AGN}},\delta_{\mathrm{AGN}}$) and redshift $z_{\mathrm{AGN}}$. Taking into account the environmental effects of BBH motion within the AGN disk, including the relativistic redshift and gravitational redshift, the effective primary mass and luminosity distance are given by:
\begin{eqnarray}
  M_{1}^{\mathrm{eff}} = & (1+z_{\mathrm{AGN}})(1+z_{\mathrm{rel}})(1+z_{\mathrm{grav}})M_{1}, \\
  D_{L}^{\mathrm{eff}} = &(1+z_{\mathrm{rel}})^{2}(1+z_{\mathrm{grav}})D_{L}(z_{\mathrm{AGN}},\Omega).
\end{eqnarray}
Here, $\Omega$ represents the cosmological parameters, which we adopt from the results reported in \citet{aghanimPlanck2018Results2020}. Under the assumption that the SMBH is non-spinning, these quantities can be expressed as:
\begin{eqnarray}
  z_{\mathrm{rel}} & = & \gamma (1+v\cos(\theta))-1, \\
  z_{\mathrm{grav}} & = & \sqrt{1-\frac{R_{\mathrm{s}}}{r}} - 1,\\
  v &=& \frac{1}{\sqrt{2(r/R_{\mathrm{s}}-1)}}.
\end{eqnarray}
Here, $\gamma = (1-v^{2})^{-1/2}$ is the Lorentz factor, $R_{s}=2GM_{\mathrm{SMBH}}/c^{2}$ is the Schwarzschild radius, $r$ is the distance between the BBH and the SMBH, $\theta$ is the viewing angle between the velocity and the observer's line of sight and $v$ is the velocity of the BBH. The final form of the evidence for the association model is thus given by:
\begin{equation}
  \label{eq:evidenceA}
 p(d|\mathcal{H}_{A}) = \int\frac{p\big(M_{1}^{\mathrm{eff}}(M_{1},r,\theta,z_{\mathrm{AGN}}),D_{L}^{\mathrm{eff}}(r,\theta,z_{\mathrm{AGN}},\Omega),\alpha_{\mathrm{AGN}},\delta_{\mathrm{AGN}}\big)}{p\big(D_{L}^{\mathrm{eff}}(r,\theta,z_{\mathrm{AGN}},\Omega)\big)}\times p(r|\mathcal{H}_{A})p(\theta|\mathcal{H}_{A})p(M_{1}|\mathcal{H}_{A})\mathrm{d}r\mathrm{d}\theta\mathrm{d}M_{1}~.
\end{equation}

We adopt a simple prior $p(r|\mathcal{H}_{A}) \propto r$ under the assumption of uniform surface density in the AGN disk, which is an axisymmetric system. The kick velocity for each remnant, as presented in Table \ref{table:GW}, is too small to have significant impacts on the orbital period. Assuming that the kicked remnant does not return to the AGN disk before 2024 October 31, we can estimate the lower bound for $r$:
\begin{equation}
    \label{eq:r}
  r > \left(\frac{t_{\mathrm{int}}}{1.6 \mathrm{yr}}\frac{10^{8}M_{\odot}}{M_{\mathrm{SMBH}}}\right)^{2/3} \times 10^{3} R_{\mathrm{s}},
\end{equation}
where $t_{\mathrm{int}}$ is the time interval between the peak time of the flare and 2024 October 31. The upper bound for $r$ is set to be $3\times 10^{3} R_{\mathrm{s}}$. The viewing angle prior $p(\theta|\mathcal{H}_{A})$ is assumed to be uniform in $\cos(\theta)$. The primary mass prior $p(M_{1}|\mathcal{H}_{A})$ is based on the mass distribution proposed by \citet{vaccaroImpactGasHardening2024} for the formation of BBH mergers in AGNs.

In the coincidence model $\mathcal{H}_{c}$, the GW event is not assumed to originate from an AGN, and the appearance of the AGN flare is regarded as a random coincidence. There are no environmental effects in this model, and the sky position of the GW event is not fixed. Thus, the evidence for the coincidence model is given by:
\begin{equation}
 p(d|\mathcal{H}_{C}) = \int \frac{p\big((1+z)M_{1},D_{L}(z,\Omega),\alpha,\delta\big)}{p\big(D_{L}(z,\Omega)\big)} \times p(\alpha| \mathcal{H}_{C})p(\delta|\mathcal{H}_{C})p(z|\mathcal{H}_{C})p(M_{1}|\mathcal{H}_{C})\mathrm{d}\alpha\mathrm{d}\delta\mathrm{d}z\mathrm{d}M_{1}~.
\end{equation}

\added{In our analysis, we implement two types of prior distributions. The default approach adopts uniform priors $p(\alpha|\mathcal{H}_{C})$ and $p(\delta|\mathcal{H}_{C})$ for sky location, and a redshift prior proportional to the comoving volume, $p(z|\mathcal{H}_{C}) \propto \frac{\mathrm{d}V_{c}}{\mathrm{d}z}$. To better account for the non-uniform spatial distribution of observed AGNs, we construct two-dimensional prior $p(\alpha,\delta|\mathcal{H}_{C})$ and three-dimensional prior $p(\alpha,\delta,z|\mathcal{H}_{C})$ based on the distribution of AGN in the Million Quasars Catalog. These priors assign higher probabilities to sky regions with greater AGN number densities. In such regions, the likelihood of chance spatial coincidences with a GW localization is naturally higher, leading to lower Bayes factors and providing a more conservative estimate of the physical association.}

\begin{figure}[htb]
  \plotone{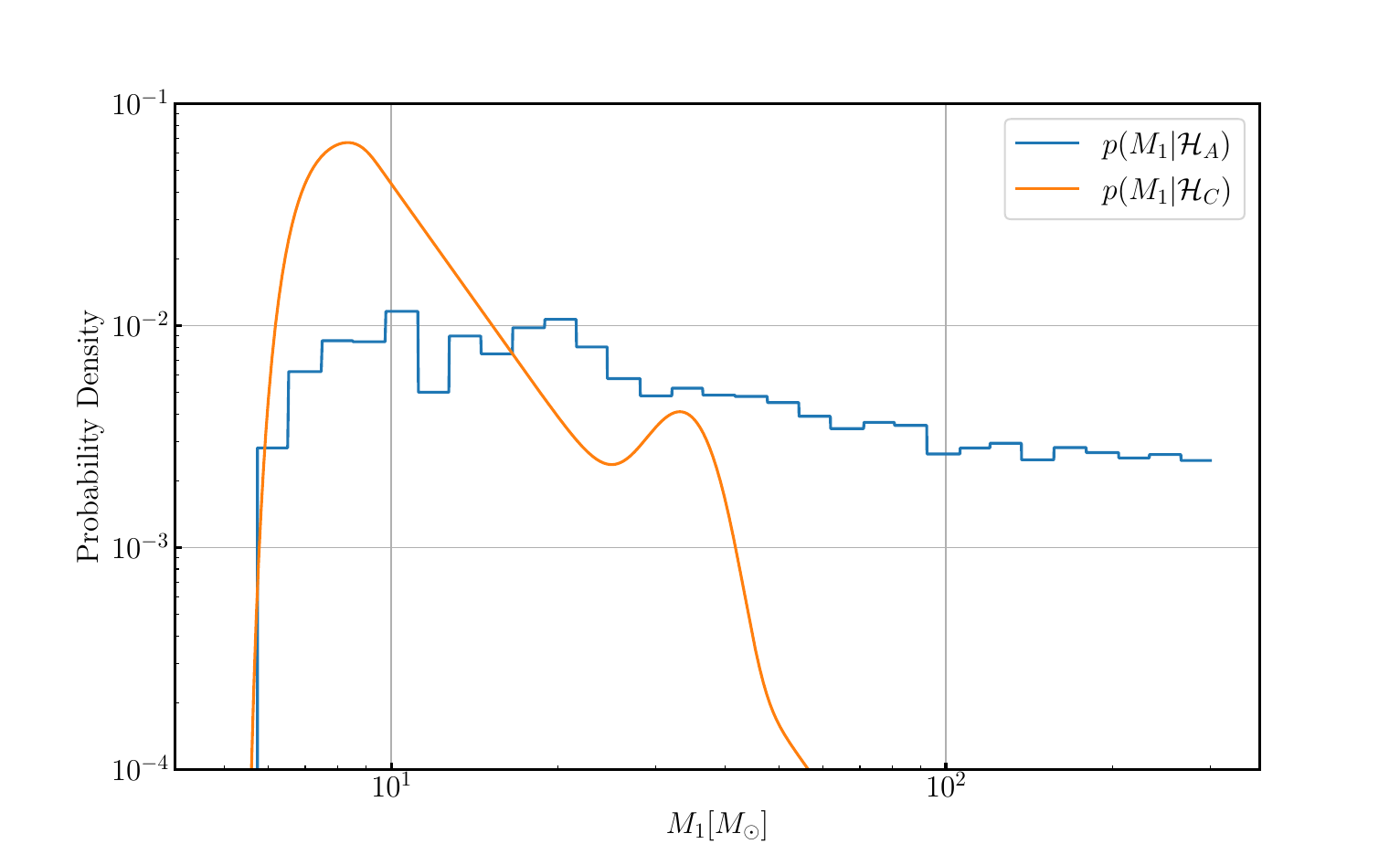}
  \caption{The primary mass prior distributions for association model (blue curve) and coincidence model (orange curve).}
  \label{fig:PrimaryMass}
\end{figure}

For the prior of the primary mass $p(M_{1}|\mathcal{H}_{C})$, we assume the \textsc{PowerLaw+Peak} model with the median parameters calculated by \citet{abbottPopulationMergingCompact2023}, which is obtained from all GW events. Figure \ref{fig:PrimaryMass} illustrates the prior distributions of the primary mass under two different models. The prior $p(M_{1}|\mathcal{H}_{A})$ is significantly higher at the high-mass end compared to $p(M_{1}|\mathcal{H}_{C})$, indicating that high-mass BBHs are more likely to form within AGN disks, thus increasing the probability of association with a specific AGN.

\section{results\label{sec:results}}

The Bayes factor, prior odds and odds ratio for each matched pair of AGN flares and GW events are presented in Table \ref{table:odds}. In general, an odds ratio greater than one ($\log\mathcal{O}^{A}_{C} > 0$) is considered to favor the association model. In this work, we interpret $\log\mathcal{O}^{A}_{C}$ values greater than 3 as positive evidence, and values exceeding 5 as very strong evidence \citep{kassBayesFactors1995a}.

\begin{deluxetable}{cccccccc}[htb]
  \tablecaption{Bayes factor ($\mathcal{B}^{A}_{C}$), prior odds ($\mathcal{P^{A}_{C}}$) and odds ratio ($\mathcal{O}^{A}_{C}$) for each matched pair of AGN flare and GW event \label{table:odds}}
  \tablehead{
   \colhead{Pair} & \colhead{AGN name} & \colhead{GW ID} & \colhead{$\log \mathcal{B}^{A}_{C}$} & \colhead{$\log \mathcal{P}^{A}_{C}$} & \colhead{$\log \mathcal{O}^{A}_{C}$} & \colhead{\added{$\log \mathcal{O}^{A}_{C}$(2D)}} & \colhead{\added{$\log \mathcal{O}^{A}_{C}$(3D)}}
  }
  \startdata
  1& J124942.30+344928.9 & GW190403\_051519 & 5.5 & -2.2 & 3.3 & \added{1.8} & \added{1.8}\\
  2& J124942.30+344928.9 & GW190514\_065416 & 2.7 & -1.1 & 1.5 & \added{0.2} & \added{-0.2}\\
  3& J124942.30+344928.9 & GW190521 & 8.8 & -0.5 & 8.3 & \added{7.0} & \added{6.5}\\
  4& J120437.98+500024.0 & GW190803\_022701 & 5.7 & 0.3 & 6.0 & \added{4.9} & \added{4.9}\\
  5& J120437.98+500024.0 & GW190909\_114149 & 3.7 & -1.4 & 2.3 & \added{2.7} & \added{2.2}\\
  6& J183412.42+365655.3 & GW190403\_051519 & 1.9 & -4.0 & -2.1 & \added{-3.5} & \added{-3.6}\\
  7& J053408.41+085450.6 & GW190731\_140936 & 4.7 & -1.2 & 3.5 & \added{4.1} & \added{3.6}\\
  8& J053408.41+085450.6 & GW190803\_022701 & 5.9 & -0.4 & 5.5 & \added{4.5} & \added{4.5}\\
  9& J181719.94+541910.0 & GW190424\_180648 & 2.5 & -11.1 & -8.5 & \added{-8.7} & \added{-9.1}\\
  10& J224333.95+760619.2 & GW190514\_065416 & 3.9 & -10.1 & -6.2 & \added{-7.5} & \added{-7.9}\\
  11& J154342.46+461233.4 & GW200216\_220804 & 5.1 & -10.4 & -5.3 & \added{-5.3} & \added{-5.8}\\
  12& J154342.46+461233.4 & GW200220\_124850 & 5.8 & -10.3 & -4.5 & \added{-5.2} & \added{-5.7}
  \enddata
  \tablecomments{\added{The last two columns show the odds ratios computed using spatially non-uniform priors derived from the two-dimensional and three-dimensional distribution of AGN in the Million Quasars Catalog. }}
\end{deluxetable}

\subsection{J124942.30+344928.9}
Our analysis identifies J124942.30+344928.9 and GW190521 as the most robust association pair, yielding an odds ratio of $\log \mathcal{O}^{A}_{C}=8.6$ \added{under the coincidence model with uniform priors}, which is consistent with the prior expectation that this pair is the most promising candidate for a common origin \citep{grahamCandidateElectromagneticCounterpart2020}. Our calculated Bayes factor ($\log \mathcal{B}^{A}_{C}=8.8$) aligns closely with the value $\log \mathcal{B}^{A}_{C} = 8.6$ reported in \citet{mortonGW190521BinaryBlack2023a}. Due to the selection of different prior odds, the odds ratio yields different results. \added{When spatially non-uniform priors derived from the AGN distribution are used instead, the odds ratio decreases, primarily due to the relatively high density of AGNs within the localization region of GW190521. Nevertheless, the odds ratio remains high, supporting a significant physical association between the AGN and this GW event.}

Additionally, two other GW events, GW190403\_051519 and GW190542\_065416, exhibit spatial and temporal coincidence with this AGN. \added{Under the coincidence model with uniform priors, the former shows positive evidence ($\log\mathcal{O}^{A}_{C} =3.3$) for association, while the latter provides only weak support and is not considered physically connected to the AGN. However, when incorporating spatially non-uniform priors derived from the AGN distribution, the evidence for GW190403\_051519 is reduced to a low level, suggesting that the initial support may be largely due to the high AGN number density in that region rather than a genuine physical association.}

Both GW190521 and GW190403\_051519 have sources with a large total mass ($\gtrsim 100 M_{\odot}$, Table \ref{table:GW}) and their primary components likely reside within the pair-instability mass gap \citep[$50-130 M_{\odot}$,][]{woosleyPulsationalPairinstabilitySupernovae2017}. This disfavors isolated binary evolution, instead supporting a dynamical formation channel like a hierarchical merger of smaller BHs. For GW190403\_051519, the high primary spin ($\chi_{1}=0.89^{+0.09}_{-0.31}$) further corroborates its second-generation origin, which is consistent with its large mass \citep{gerosaHierarchicalMergersStellarmass2021}. The low mass ratio ($q=0.23^{+0.57}_{-0.12}$) of GW190403\_051519 is unusual in isolated binary evolution \citep{marchantNewRouteMerging2016,eldridgeBpassPredictionsBinary2016}, but it is expected if the primary and secondary components are a second- and a first-generation BH, respectively \citep{rodriguezBlackHolesNext2019}. All these features hint that these two BBHs were more likely to originate from dense environments.

The absence of a secondary flare in J124942.30+344928.9 appears inconsistent with earlier predictions of recurrent activity \citep{grahamCandidateElectromagneticCounterpart2020}. This discrepancy arises from the dependence of the recurrence timescale ($t_{\mathrm{rec}}$) on both the central SMBH mass and the distance between BBH and SMBH, following the relation $t_{\mathrm{rec}} \propto M_{\mathrm{SMBH}}(r/R_{s})^{3/2} $. While current estimates of $M_{\mathrm{SMBH}}$ carry significant uncertainties, even assuming that the measurement in \citet{grahamLightDarkSearching2023} represents the true mass ($\sim 10^{8.6} M_{\odot}$), the unknown value of $r$ prevents definitive constraints. Applying Eq. \ref{eq:r} to this system yields $r \gtrsim 900 R_{\mathrm{s}}$, corresponding to approximately 0.03 pc for the adopted $M_{\mathrm{SMBH}}$. Continued monitoring of this AGN is essential to probe the recurrence window and elucidate the physical origin of its transient signal.

\subsection{J120437.98+500024.0}
Another notable association arises between AGN J120437.98+500024.0 and GW190803\_022701, yielding a high odds ratio of $\log \mathcal{O}^{A}_{C}=6.0$ \added{under the coincidence model with uniform priors, and still substantially value of $\log \mathcal{O}^{A}_{C}=4.9$ with both 2D and 3D non-uniform priors}. Although the primary mass ($M_{1}=38^{+10}_{-7} M_{\odot}$) is not particularly high, its well-constrained sky area ($A_{90}=1021 \  \mathrm{deg}^{2}$) substantially enhances the positional coincidence probability. This small localization area contains relatively few AGNs, reducing the potential for chance coincidence. When coupled with the significance of the flare in this AGN, a high prior odds is obtained, leading to a high level of this association. In contrast, GW190909\_114149 shows weaker evidence for association, so we exclude it from further consideration.

\subsection{J183412.42+365655.3}
The other AGN with a high-confidence flare, J183412.42+365655.3, is possibly associated with GW190403\_051519 in \citet{grahamLightDarkSearching2023}. However, this source lies at the edge of the 90\% credible sky localization region of this GW event, resulting in a small Bayes factor and a low odds ratio. This value renders the association statistically unreliable, suggesting the observed flare originates from other astrophysical processes rather than the BBH merger.

\subsection{Others}
While the remaining four AGNs are excluded in Section \ref{sec:pflare} through light curve analysis, we compute their odds ratio for completeness. Notably, AGN J053408.41+085450.6 exhibits positive evidence for association with GW190731\_140936 and GW190803\_022701, and there is a recurrent flare approximately three years after the initial one in this AGN, with the peak luminosity comparable to the earlier flare. If this AGN is not a blazar, it would represent a compelling optical counterpart for BBH mergers.

For the other three AGNs (J181719.94+541910.0, J224333.95+760619.2 and J154342.46+461233.4), the low odds ratios ($\log \mathcal{O}^{A}_{C} < 0$) decisively favor the coincidence hypothesis. This outcome is driven by their low flare significance, which directly suppresses the prior odds and consequently the overall odds ratio. The consistency between low $p_{\mathrm{flare}}$ and low odds ratio underscores the effectiveness of our filtering approach in prioritizing robust candidates.

\section{Constraining the Hubble Constant\label{sec:hubble}}
The Hubble constant $H_{0}$ is a fundamental parameter in cosmology that describes the rate at which the universe is expanding. By associating GW events with EM counterparts, one can estimate the distance and redshift independently, and the joint observation of a source can be used to constrain $H_{0}$ and potentially ease the Hubble tension \citep{divalentinoRealmHubbleTension2021}. Currently, GW170817 \citep{abbottGW170817ObservationGravitational2017} is the only GW event with a confirmed EM counterpart \citep{goldsteinOrdinaryShortGammaRay2017,savchenkoINTEGRALDetectionFirst2017}, and its joint observation yields a value of $H_{0}=70.0^{+12.0}_{-8.0}\mathrm{km \ s^{-1} Mpc^{-1}}$ \citep{abbottGravitationalwaveStandardSiren2017}.

$H_{0}$ has also been estimated based on the association between GW190521 and J124942.30+344928.9 \citep{mukherjeeFirstMeasurementHubble2020,gayathriMeasuringHubbleConstant2021,chenStandardSirenCosmological2022,mortonGW190521BinaryBlack2023a}. While the results vary slightly, they are consistent with those derived from Planck \citep{aghanimPlanck2018Results2020} and SH0ES \citep{riessComprehensiveMeasurementLocal2022}. Based on our findings, several other pairs of AGNs and GW events may also be associated. According to these associations, we estimate $H_{0}$ following the method outlined in \citet{mortonGW190521BinaryBlack2023a}.

To calculate the evidence for the association model with Eq. \ref{eq:evidenceA}, we make use of the cosmological parameters $\Omega$. Instead of fixing the whole set of cosmological parameters, we treat $H_{0}$ as a free parameter while keeping other parameters (density and dark energy equation of state parameters) unchanged, thus we can derive the posterior distribution of $H_{0}$. We employ two types of priors: a flat prior, where $H_{0}$ is uniformly distributed within the range [20,160], and a prior based on the results of GW170817 \citep{abbottGravitationalwaveStandardSiren2017}. Using these priors, we estimate $H_{0}$ for multiple AGN-GW pairs.

\begin{figure}
  \centering
  \gridline{
    \includegraphics[trim=30 10 30 50, clip, width=0.5\textwidth]{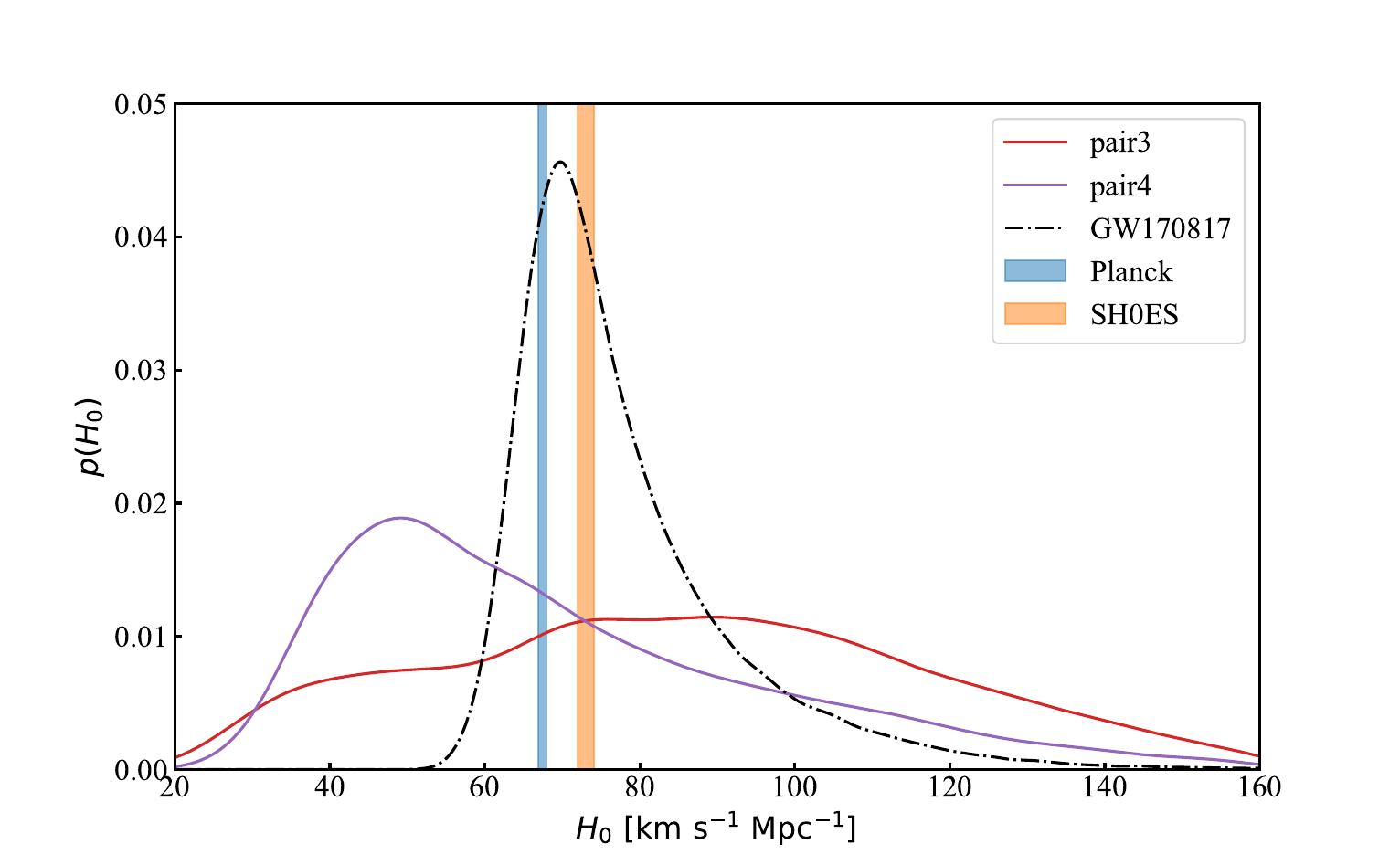}
    \includegraphics[trim=30 10 30 50, clip, width=0.5\textwidth]{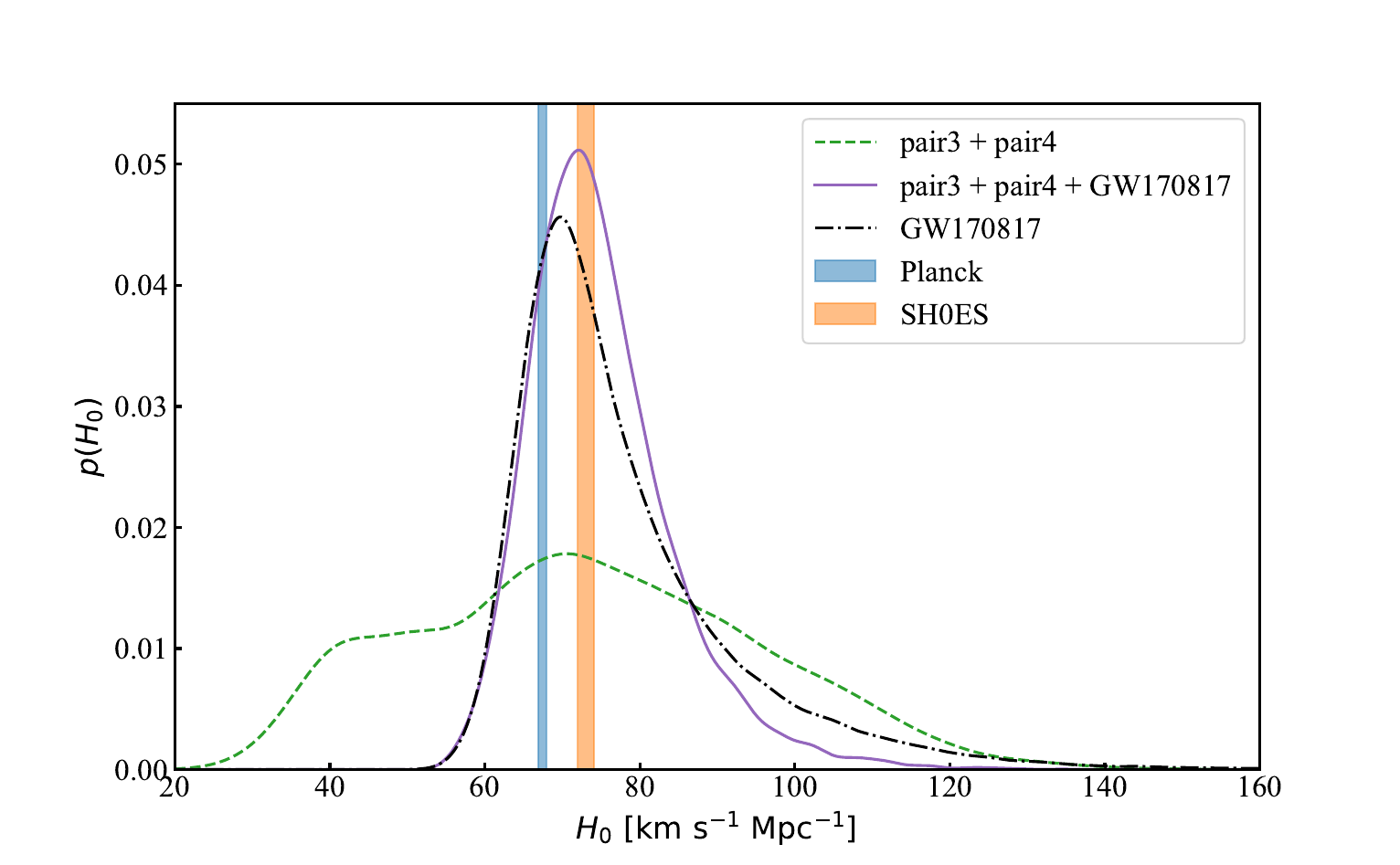}
  }
  \caption{Hubble constant posterior distributions for single pair (left) and pair combination (right), compared to the reported values from Planck and SH0ES (1-$\sigma $ credible interval).  }
  \label{fig:H0}
\end{figure}

\added{Figure \ref{fig:H0} shows the $H_{0}$ posterior distribution estimated from two potentially associated pairs, both individually and in combination. The distributions are consistent with the results from Planck and SH0ES but exhibit large uncertainties. When combining two pairs, the uncertainty is reduced, yielding $H_{0}=72.1^{+23.9}_{-23.1}\ \mathrm{km \ s^{-1} Mpc^{-1}}$ under the flat prior. Using the GW170817 prior instead results in a tighter constraint of $H_{0}=73.5^{+9.8}_{-6.9}\  \mathrm{km \ s^{-1} Mpc^{-1}}$. A summary of all $H_{0}$ estimation is provided in Table \ref{table:H0}.}

\begin{deluxetable}{ccc}[htb]
  \tablecaption{Hubble constant estimated for different pairs with different priors (uncertainties represent the 68\% credible interval) \label{table:H0}}
  \tablehead{
    \colhead{Pair(s)} & \colhead{$H_{0}$ (Uniform prior)} & \colhead{$H_{0}$ (GW170817 prior)} \\
    & \colhead{$\mathrm{km \ s^{-1} Mpc^{-1}}$} & \colhead{$\mathrm{km \ s^{-1} Mpc^{-1}}$}}
  \startdata
  3 & $85.4^{+33.6}_{-35.2}$ & $75.2^{+13.7}_{-8.6}$ \\
  4 & $62.4^{+35.7}_{-19.1}$ & $69.7^{+9.2}_{-5.6}$ \\
  3+4 & $72.1^{+23.9}_{-23.1}$ & $73.5^{+9.8}_{-6.9}$\\
  \enddata
\end{deluxetable}

\section{Discussions and conclusions\label{sec:discussion}}
We present a systematic re-evaluation of seven AGN flares proposed as potential electromagnetic counterparts to GW events detected by LVK collaboration \citep{grahamLightDarkSearching2023} with three additional years of ZTF public data. While the original analysis reported that these AGNs show the probability of having a flare exceeding 99.5\% in both g- and r-band, our extended temporal baseline reveals that three flares are statistically consistent with stochastic AGN variability rather than transient merger signals. Furthermore, the classification of J053408.41+085450.6 as a likely blazar eliminates its candidacy, leaving only three AGNs possible to be associated with GW events.

We quantify the associations between AGN flares and GW events using the Bayesian statistical framework from \citet{mortonGW190521BinaryBlack2023a}. First, we estimate the AGN number within the 90\% credible volume of each GW event using the Million Quasar Catalog, then compute prior odds for each AGN-GW pair by combining the number with flare occurrence probabilities in individual AGN. Subsequently, we evaluate the Bayes factor for two competing hypotheses: common origin and chance coincidence. The key factors influencing this calculation are the location of the AGN and the primary mass of the GW event. \added{To better account for the non-uniform spatial distribution of AGNs, we also consider non-uniform priors derived from the distribution of AGNs in the Million Quasars Catalog under the coincidence model, which modify the Bayes factor depending on the local AGN density within the GW localization region.} Finally, by combining the prior odds and Bayes factor to calculate the odds ratio, we find that two AGN-GW pairs, pair 3 (J124942.30+344928.9 and GW190521) and pair 4 (J120437.98+500024.0 and GW190803\_022701), show a positive tendency for association.

We conclude that only two flares, J124942.30+344928.9 and J120437.98+500024.0, could be associated with GW events, while the remaining flares are excluded for various reasons. However, neither of the AGNs hosting these two flares shows secondary flaring activity until 2024 October 31, which is predicted to occur after several years \citep{grahamCandidateElectromagneticCounterpart2020}.

We estimate the Hubble constant for these potential AGN-GW pairs with different priors. All of the results are consistent with both Planck \citep{aghanimPlanck2018Results2020} and SH0ES \citep{riessComprehensiveMeasurementLocal2022} measurements. For the most promising AGN-GW pairs combination, we report $H_{0}=72.1^{+23.9}_{-23.1}\ \mathrm{km \ s^{-1} Mpc^{-1}}$ with a flat prior and  $H_{0}=73.5^{+9.8}_{-6.9}\  \mathrm{km \ s^{-1} Mpc^{-1}}$ with GW170817 prior, hint towards a larger value of $H_{0}$.

Continued monitoring of the two AGNs identified as potential EM counterparts is essential. In particular, it is important to track these AGNs for the occurrence of a secondary flare. If such a flare is detected, prompt spectroscopic observations must be carried out to provide critical insight that could definitely establish or rule out the association between these flares and the corresponding GW events. Moreover, the significantly improved GW localization during O4 will greatly enhance our ability to search for EM counterparts. The discovery of additional EM counterparts in the future will not only refine our understanding of the associations between AGN and GW events but also contribute to a deeper insight into the underlying astrophysical processes and the evolution of the universe.


\begin{acknowledgements}

We appreciate the helpful discussion with Tinggui Wang, Jianmin Wang and Lian Tao. This work is supported by Strategic Priority Research Program of the Chinese Academy of Science (Grant No. XDB0550300), the National Key R\&D Program of China (Grant No. 2021YFC2203102 and 2022YFC2204602), the National Natural Science Foundation of China (Grant No. 12325301 and 12273035), the Science Research Grants from the China Manned Space Project (Grant No. CMS-CSST-2021-B01), the 111 Project for ``Observational and Theoretical Research on Dark Matter and Dark Energy" (Grant No. B23042), and Cyrus Chun Ying Tang Foundations. 

This research has made use of data or software obtained from the Gravitational Wave Open Science Center (gwosc.org), a service of the LIGO Scientific Collaboration, the Virgo Collaboration, and KAGRA. This material is based upon work supported by NSF's LIGO Laboratory which is a major facility fully funded by the National Science Foundation, as well as the Science and Technology Facilities Council (STFC) of the United Kingdom, the Max-Planck-Society (MPS), and the State of Niedersachsen/Germany for support of the construction of Advanced LIGO and construction and operation of the GEO600 detector. Additional support for Advanced LIGO was provided by the Australian Research Council. Virgo is funded, through the European Gravitational Observatory (EGO), by the French Centre National de Recherche Scientifique (CNRS), the Italian Istituto Nazionale di Fisica Nucleare (INFN) and the Dutch Nikhef, with contributions by institutions from Belgium, Germany, Greece, Hungary, Ireland, Japan, Monaco, Poland, Portugal, Spain. KAGRA is supported by Ministry of Education, Culture, Sports, Science and Technology (MEXT), Japan Society for the Promotion of Science (JSPS) in Japan; National Research Foundation (NRF) and Ministry of Science and ICT (MSIT) in Korea; Academia Sinica (AS) and National Science and Technology Council (NSTC) in Taiwan. This work is based on observations obtained with the Samuel Oschin Telescope 48-inch and the 60-inch Telescope at the Palomar Observatory as part of the Zwicky Transient Facility project. ZTF is supported by the National Science Foundation under Grants No. AST-1440341 and AST-2034437 and a collaboration including current partners Caltech, IPAC, the Oskar Klein Center at Stockholm University, the University of Maryland, University of California, Berkeley , the University of Wisconsin at Milwaukee, University of Warwick, Ruhr University, Cornell University, Northwestern University and Drexel University. Operations are conducted by COO, IPAC, and UW.
\end{acknowledgements}

\software{ astropy \citep{astropy:2013,astropy:2018,astropy:2022}, matplotlib \citep{Hunter:2007}, numpy \citep{harris2020array}, FIGARO \citep{rinaldiRapidLocalizationGravitational2022}, celerite \citep{foreman-mackeyFastScalableGaussian2017}, ligo.skymap \citep{singerRapidBayesianPosition2016}, surfinBH \citep{https://doi.org/10.5281/zenodo.1435832}.
}

\bibliography{main}
\bibliographystyle{aasjournalv7}
\end{document}